\documentclass[%
superscriptaddress,
preprint,
showpacs,
nofootinbib,
 amsmath,amssymb,
 aps,   
prd,
]{revtex4-2}

\usepackage{amsmath}
\usepackage{amsfonts} 
\usepackage{mathtools} 
\usepackage{amssymb} 
\usepackage{subfig} 
\usepackage{graphicx}
\usepackage{xcolor}
\usepackage{mathrsfs}
\usepackage[breaklinks,colorlinks,urlcolor=blue,citecolor=blue,linkcolor=magenta]{hyperref}
\usepackage{multirow}
\usepackage{orcidlink}
\usepackage{cleveref}
\crefname{table}{Table}{Tables}
\crefname{equation}{Eq.}{Eqs.}
\crefname{appendix}{App.}{Apps.}
\crefname{section}{Sec.}{Secs.}
\crefname{figure}{Fig.}{Figs.}

\usepackage{etoolbox}

\newcommand{\refEq}[1]{Eq.\,\eqref{#1}} 
\newcommand{\refEqs}[1]{Eqs.\,\eqref{#1}} 
\newcommand{\refFig}[1]{Fig.\,\ref{#1}}
\newcommand{\refFigs}[1]{Figs.\,\ref{#1}}
\newcommand{\abs}[1]{|#1|} 
\newcommand{\re}[1]{\text{Re}\left(#1\right)}
\newcommand{\im}[1]{\text{Im}\left(#1\right)}

\newcommand{\SD}[1]{\Phi_{#1}^{\phantom{\dagger}}}
\newcommand{\SDd}[1]{\Phi_{#1}^\dagger}
\newcommand{\SDc}[1]{\Phi_{#1}^\ast}

\newcommand{\bilH}[1]{\mathcal H_{#1}}
\newcommand{\bilA}[1]{\mathcal A_{#1}}

\newcommand{\qq}[1]{\mu^2_{#1}}
\newcommand{\QQ}[1]{\lambda_{#1}}
\newcommand{\QQA}[1]{\lambda^{\mathcal{A}}_{#1}}
\newcommand{\co}[2]{c_{#1[#2]}}
\newcommand{\si}[2]{s_{#1[#2]}}

\newcommand{\nfR}[1]{\mathrm{R}_{#1}}
\newcommand{\nfI}[1]{\mathrm{I}_{#1}}
\newcommand{\chfPM}[1]{\mathrm{C}^\pm_{#1}}
\newcommand{\chfP}[1]{\mathrm{C}^+_{#1}}
\newcommand{\chfM}[1]{\mathrm{C}^-_{#1}}
\newcommand{\chfvec}{{\vec{\mathbf{C}}}}
\newcommand{\nfvec}{{\vec{\mathbf{N}}}}
\newcommand{\CHfvec}{{\vec{\mathbf{H}}_-}}
\newcommand{\Nfvec}{{\vec{\mathbf{H}}_0}}
\newcommand{\FchfPM}[1]{\mathrm{H}^\pm_{#1}}
\newcommand{\FchfP}[1]{\mathrm{H}^+_{#1}}
\newcommand{\FchfM}[1]{\mathrm{H}^-_{#1}}
\newcommand{\Fnf}[1]{\mathrm{H}_{#1}^0}
\newcommand{\Omat}{O}\newcommand{\OmatT}{\Omat^T}
\newcommand{\Omatel}[1]{\Omat_{#1}}
\newcommand{\Umat}{U}\newcommand{\UmatDag}{\Umat^\dagger}

\newcommand{\Umatcel}[1]{\Umat_{#1}^\ast}

\newcommand{\VEV}[1]{\langle #1 \rangle}
\newcommand{\vev}[1]{v_{#1}}
\newcommand{\vevPh}[1]{\theta_{#1}}

\newcommand{\nMM}{M_0^2}

\newcommand{\chMM}{M_{\pm}^2}

\newcommand{\potV}{\mathcal{V}}
\newcommand{\vevV}{\mathrm{V}}
\newcommand{\TR}[1]{\text{Tr}\left(#1\right)}

\def\eg{\textit{e.g.}}
\def\ie{\textit{i.e.}}

\graphicspath{{Figs/}}

\makeatletter
\def\frontmatter@abstractheading{\vspace{0.8cm}}
\makeatother

\begin{document}

\title{Light states in real 3HDMs with spontaneous CP violation and softly broken symmetries}

\author{José M. Camacho\orcidlink{0009-0001-8856-258X}}\email{jose.m.camacho@uv.es}
\affiliation{%
Departament de Física Teòrica \& Instituto de Física Corpuscular (IFIC),\\ Universitat de València -- CSIC, E-46100 Valencia, Spain
}
\author{Carlos Miró\orcidlink{0000-0003-0336-9025}}\email{carlos.miroarenas@to.infn.it}
\affiliation{%
INFN, Sezione di Torino,\\ Via Pietro Giuria 1,
I-10125 Turin, Italy
}
\author{Miguel Nebot\orcidlink{0000-0001-9292-7855}}\email{Miguel.Nebot@uv.es}
\author{Daniel Queiroz\orcidlink{0009-0000-0554-0530}}\email{Daniel.Queiroz@uv.es}
\author{Tomás Tobarra\orcidlink{0009-0000-6773-4023}}\email{Tomas.Tobarra@ific.uv.es}
\affiliation{%
Departament de Física Teòrica \& Instituto de Física Corpuscular (IFIC),\\ Universitat de València -- CSIC, E-46100 Valencia, Spain
}

\begin{abstract}
Scalar sectors with several Higgs doublets, a CP invariant potential, and a CP violating vacuum, possess a mass spectrum in which, unexpectedly, new scalars cannot have masses much larger than the electroweak scale once perturbativity requirements are imposed on the quartic couplings of the Higgs potential and despite the presence of free quadratic (mass) terms that can be arbitrarily large. The minimal model in which this behavior is manifest involves 3 Higgs doublets. We analyze and illustrate in detail that kind of scenario including an additional simplifying assumption: the quartic part of the potential is shaped by some discrete symmetry. Besides analytic results, a numerical analysis is presented together with some phenomenological consequences derived only from the properties of the scalar sector alone. 
\end{abstract}

\maketitle

\clearpage
\section{Introduction\label{SEC:Intro}}
In any extension of the Standard Model (SM), knowledge of the masses of the new particles is of utmost importance. Besides direct experimental evidence or indirect hints, different theoretical arguments can be invoked in order to set some kind of constraint on the new particle spectra. A remarkable historical example is found on the hunt for the SM Higgs boson, where perturbativity requirements allowed to set bounds on its mass well before its discovery \cite{ATLAS:2012yve,CMS:2012qbp}, as proposed by B. Lee, Quigg and Thacker in \cite{Lee:1977eg,Lee:1977yc} --- see also \cite{Dicus:1973gbw}. To this aim, other theoretical avenues were explored, for example, in \cite{Weinberg:1976pe,Politzer:1978ic,Cabibbo:1979ay,Dashen:1983ts,Callaway:1983zd}. In the context of extended scalar sectors featuring spontaneous electroweak symmetry breaking, perturbativity constraints imply the existence of a Higgs-like state whose mass cannot be much larger than the symmetry breaking vacuum expectation value (vev) --- the electroweak scale $\vev{}\simeq 246$ GeV ---, as discussed in \cite{Weldon:1984th,Langacker:1984dma} (see also \cite{Comelli:1996xg,Espinosa:1996vt}).

Narrowing down on multi-Higgs models \cite{Ivanov:2017dad}, in the minimal case with two Higgs doublets (2HDMs) --- first proposed by T.D. Lee \cite{Lee:1973iz,Lee:1974jb} ---, and in particular when some symmetry is imposed, perturbativity related constraints on the masses of the new scalars have been broadly explored in the literature, \eg\  \cite{Huffel:1980sk,Casalbuoni:1987cz,Maalampi:1991fb,Kanemura:1993hm,Ginzburg:2005dt,Horejsi:2005da,Kanemura:2015ska}. The extreme case, not exclusive of multi-Higgs scenarios, is a scalar sector where a symmetry requirement only allows some quadratic parameters and, through stationarity conditions, they can be entirely expressed in terms of dimensionless quartic parameters and powers of vacuum expectation values. Then, all the new scalars have bounded masses. Free unconstrained quadratic parameters are certainly necessary if a regime is to be achieved in which some of the new scalars are arbitrarily heavy. Different aspects of such decoupling regimes are discussed, for example, in \cite{Haber:1989xc,Bhattacharyya:2014oka,Faro:2020qyp,Carrolo:2021euy}; for softly broken discrete symmetries, see \cite{deMedeirosVarzielas:2021zqs}.

In multi-Higgs models consisting of $n$ Higgs doublets with a CP invariant scalar potential --- hereafter dubbed \emph{real} $n$HDMs --- and spontaneous CP violation (SCPV), it was discussed in \cite{Miro:2024zka} that, among the new scalars, several states --- one charged and two neutral ones, in addition to ``the Higgs'' ---, cannot have masses much larger than the electroweak scale if quartic couplings in the potential obey perturbativity constraints. This unexpected property holds no matter how many independent quadratic parameters are left unconstrained by the stationarity conditions. In the 2HDM case, there are no independent quadratic parameters left unconstrained by the stationarity conditions, and that outcome is thus straightforward: perturbativity constraints bound the whole scalar spectrum \cite{Nebot:2018nqn,Nebot:2019qvr,Nierste:2019fbx}. With more than 2 Higgs doublets, that outcome is however non-trivial: the number of free quadratic parameters increases in fact quadratically with the number of Higgs doublets and one could have expected that a regime with an arbitrary large gap between the electroweak scale and the masses of all new scalars is possible. Surprisingly, that is not the case. As summarized for completeness in \cref{SEC:rnHDMSCPV}, the key point is the following: in addition to the chosen (spontaneous) CP violating vacuum, its CP conjugate is also a good candidate vacuum. Considering the regime with large free quadratic parameters that might produce large masses for the new scalars, one can neglect in first approximation the quartic couplings. Then, the mass matrices around the chosen vacuum do not distinguish the CP conjugate candidate vacua, implying that the number of massless states in that regime is doubled: one additional massless state for each would-be Goldstone boson, and another additional massless state for the massless ``Higgs''. Reintroducing then the quartic couplings as a perturbation, the true would-be Goldstone bosons remain massless, while the remaining massless states (one charged and three neutral ones, including the ``Higgs'') get masses of the order of the electroweak scale --- more precisely, squared masses $M^2 \sim$ (bounded $\QQ{}$) $\times$ (vev)$^2$.

The goal of this work is to illustrate in detail this generic property in a setup that allows a more transparent analytic understanding. In terms of the number of Higgs doublets involved, the minimal choice of interest is $n=3$: in that case, the stationarity conditions leave one unconstrained quadratic parameter. The quartic part of the potential includes, nevertheless, many parameters: to retain some analytic control, we consider scalar potentials where that quartic part is shaped by some discrete symmetry. %
Specifically, we focus on cases where they transform as irreducible triplets of some discrete group such that the quartic part of the potential is invariant. Popular scenarios addressed in the literature are, for example, the groups $A_4$, $\Delta(27)$, and, with $n>3$, $\Delta(3n^2)$ and $\Delta(6n^2)$ \cite{Ma:2001dn,Ivanov:2012fp,deMedeirosVarzielas:2017ote,Kobayashi:2022moq}. Rather than the scalar sector \emph{per se}, the motivation for these choices is often related to fermion masses and mixings. Nevertheless, that is beyond the goals and scope of this work: we choose these examples because they provide a quartic part of the scalar potential with a motivated reduced number of parameters. The symmetry is not imposed on the quadratic part of the potential, that is, it is ``softly broken'', while CP invariance is imposed on the complete potential.

The manuscript is organized as follows. In \cref{SEC:rnHDMSCPV} we provide a brief overview of the general scenario discussed in \cite{Miro:2024zka}. In \cref{SEC:sym3HDMSCPV} we discuss in turn the scenarios with $A_4$ and $\Delta(27)$ symmetries --- the cases of $\Delta(3n^2)$ and $\Delta(6n^2)$ with $n>3$ are obtained from the $A_4$ analysis taking a simple parametric limit. In the $A_4$ scenario, the addition of CP invariance implies that, rather than invariance under $A_4$, one is effectively imposing invariance under $S_4$: despite this nuance, we still refer to it as the $A_4$ case in the following. Although a number of relevant analytic results are presented in \cref{SEC:sym3HDMSCPV}, a full numerical study from which some phenomenological implications of interest can be derived is presented in \cref{SEC:NumPheno}. Finally, we present our conclusions. Additional details on the numerical exploration and the $2\to 2$ perturbative unitarity constraints are provided in Apps.~\ref{APP:PotNum} and \ref{APP:2to2}, respectively.

\section{Light states in real $\boldsymbol{n}$HDMs with SCPV\label{SEC:rnHDMSCPV}}
We start with a brief overview of the main results of \cite{Miro:2024zka}. 
The most general scalar potential for $n$ Higgs doublets $\SD{a}$ ($a = 1,\dots,n$), invariant under the CP transformation $\SD{a}\mapsto\SDc{a}$, reads
\begin{equation}\label{eq:RealnHDM:V:01}
 \potV(\SD{1},\ldots,\SD{n})=\potV_2(\SD{1},\ldots,\SD{n})+\potV_4(\SD{1},\ldots,\SD{n})\,,
\end{equation}
with
\begin{equation}\label{eq:RealnHDM:V2:01}
 \potV_2(\SD{1},\ldots,\SD{n})=\sum_{a=1}^n\qq{a}\SDd{a}\SD{a}+\sum_{a=1}^{n-1}\sum_{b=a+1}^n\qq{ab}\bilH{ab}\,,
\end{equation}
and
\begin{equation}\label{eq:RealnHDM:V4:01}
 \begin{aligned}
 &\potV_4(\SD{1},\ldots,\SD{n})=
\sum_{a=1}^n\QQ{a}(\SDd{a}\SD{a})^2+\sum_{a=1}^{n-1}\sum_{b=a+1}^{n}\QQ{a,b}(\SDd{a}\SD{a})(\SDd{b}\SD{b})\\
 &+\sum_{a=1}^n\sum_{b=1}^{n-1}\sum_{c=b+1}^n\QQ{a,bc}(\SDd{a}\SD{a})\bilH{bc}
+\left.\sum_{a=1}^{n-1}\sum_{b=a+1}^n\sum_{c=1}^{n-1}\sum_{d=c+1}^n\right|_{(a,b)\leq(c,d)}\hspace{-8ex}\left(\QQ{ab,cd}\bilH{ab}\bilH{cd}+\QQA{ab,cd}\bilA{ab}\bilA{cd}\right)\,.
\end{aligned}
\end{equation}
In \cref{eq:RealnHDM:V2:01,eq:RealnHDM:V4:01}, we use the hermitian and antihermitian bilinears
\begin{equation}\label{eq:bil}
\bilH{ab}\equiv\frac{1}{2}(\SDd{a}\SD{b}+\SDd{b}\SD{a})\,,\qquad \bilA{ab}\equiv\frac{1}{2}(\SDd{a}\SD{b}-\SDd{b}\SD{a})\,,\qquad a<b\,, 
\end{equation}
which are, respectively, CP-even and CP-odd. 
All quadratic parameters $\qq{a}$, $\qq{ab}$ in $\potV_2$, and quartic parameters $\QQ{a}$, $\QQ{a,b}$, $\QQ{a,bc}$, $\QQ{ab,cd}$, $\QQA{ab,cd}$ in $\potV_4$ are real attending to CP invariance  (hence \emph{real} $n$HDM). In the last term in \cref{eq:RealnHDM:V4:01}, $(a,b)\leq(c,d)$ stands for $a\leq c$ and $b\leq d$. Spontaneous electroweak symmetry breaking occurs for a minimum of the function
\begin{equation}
 \vevV(\VEV{\SD{1}},\ldots,\VEV{\SD{n}})\equiv\potV(\VEV{\SD{1}},\ldots,\VEV{\SD{n}})\,,
\end{equation}
with $\VEV{\SD{a}}\neq\left(\begin{smallmatrix}0\\ 0\end{smallmatrix}\right)$. Appropriate electroweak spontaneous symmetry breaking is assumed for the vevs $\VEV{\SD{a}}$, and the fields are expanded as
\begin{equation}\label{eq:FieldExp:01}
 \SD{a}=\frac{e^{i\vevPh{a}}}{\sqrt 2}\begin{pmatrix}\sqrt{2}\chfP{a}\\ \vev{a}+\nfR{a}+i\,\nfI{a}\end{pmatrix},\quad \text{where}\quad \langle\SD{a}\rangle=\frac{\vev{a}e^{i\vevPh{a}}}{\sqrt 2}\begin{pmatrix}0\\ 1\end{pmatrix},\quad \vev{a},\vevPh{a}\in\mathbb{R}\,,\ \ \vev{a}\geq 0\,.
\end{equation}
The phases $\vevPh{a}$ produce spontaneous CP violation; more precisely, only phase differences $\vevPh{b}-\vevPh{a}$ (mod $2\pi$) matter --- an overall rephasing is just equivalent to a global hypercharge transformation. Stationarity conditions $\partial_{\vev{a}}\vevV=0$ and  $\partial_{\vevPh{a}}\vevV=0$ are necessary in order to have a minimum at the desired vevs. There are $2n-1$ stationarity relations since $\sum_{a=1}^n\partial_{\vevPh{a}}\vevV=0$ by construction. They have the following form:
\begin{equation}\label{eq:RealnHDM:Stationarity:v:01}
 \partial_{\vev{a}}{\vevV}=\qq{a}\vev{a}+\frac{1}{2}\sum_{b< a}\qq{ba}\co{}{ba}\vev{b}+\frac{1}{2}\sum_{b>a}\qq{ab}\co{}{ab}\vev{b}+[(\QQ{}\text{'s})\times(\text{vevs})^3]\,,
\end{equation}
and
\begin{equation}\label{eq:RealnHDM:Stationarity:th:01}
 \partial_{\vevPh{a}}{\vevV}=\frac{\vev{a}}{2}\left(\sum_{b<a}\qq{ba}\si{}{ba}\vev{b} - \sum_{b>a}\qq{ab}\si{}{ab}\vev{b}\right)+[(\QQ{}\text{'s})\times(\text{vevs})^4]\,,
\end{equation}
where $[(\QQ{}\text{'s})\times(\text{vevs})^3]$ and $[(\QQ{}\text{'s})\times(\text{vevs})^4]$ stand for terms linear in the quartic couplings $\QQ{}$ and cubic/quartic, respectively, in vacuum expectation values $\vev{a}$. In \cref{eq:RealnHDM:Stationarity:v:01,eq:RealnHDM:Stationarity:th:01}, $\si{}{ab}\equiv\sin\Theta_{ab}$ and $\co{}{ab}\equiv\cos\Theta_{ab}$, where $\Theta_{ab}\equiv\vevPh{a}-\vevPh{b}$.

It is straightforward to check that, out of the $n(n+1)/2$ quadratic parameters, only $n(n+1)/2-(2n-1)=n(n-3)/2+1$ remain unconstrained once the stationarity conditions are taken into account. For instance, one can fix $\qq{1b}$ with $b=2,\ldots,n$ using $\partial_{\vevPh{b}}\vevV=0$, and then fix $\qq{a}$ with $a=1,\ldots,n$ using $\partial_{\vev{a}}\vevV=0$, yielding all $\qq{ab}$, with $a=2,\ldots,n-1$ and $b>a$, unconstrained.

The mass terms $-\mathscr L_{\mathrm{Mass}}\subset\potV$ for charged $\chfvec^\dagger=(\chfP{1},\ldots,\chfP{n})$ and neutral fields $\nfvec^T=(\nfR{1},\ldots,\nfR{n},\nfI{a},\ldots,\nfI{n})$, are
\begin{equation}\label{eq:pot:massterms:01}
 -\mathscr L_{\mathrm{Mass}}=\chfvec^\dagger\,\chMM\,\chfvec+\frac{1}{2}\nfvec^T\nMM\nfvec\,,
\end{equation}
where $\chMM$ is the $n\times n$ hermitian charged mass matrix, and $\nMM$ is the $2n\times 2n$ real symmetric neutral mass matrix, with elements
\begin{equation}\label{eq:chMM}
(\chMM)_{a,b}=\left[\frac{\partial^2\potV}{\partial\chfP{a} \partial\chfM{b}}\right],
\end{equation}
and
\begin{equation}\label{eq:nMM}
(\nMM)_{a,b}=\left[\frac{\partial^2\potV}{\partial\nfR{a} \partial\nfR{b}}\right],\quad
(\nMM)_{a,n+b}=(\nMM)_{n+b,a}=\left[\frac{\partial^2\potV}{\partial\nfR{a} \partial\nfI{b}}\right],\quad
(\nMM)_{n+a,n+b}=\left[\frac{\partial^2\potV}{\partial\nfI{a} \partial\nfI{b}}\right].\\
\end{equation}
In \cref{eq:chMM,eq:nMM}, the second derivatives in squared brackets $[\ldots]$ are evaluated at vanishing values of the fields, namely, $\chfPM{a}$, $\nfR{a}$, $\nfI{a}\to 0$. The physical charged and neutral scalars are the eigenvectors of the mass matrices, with their squared masses the corresponding eigenvalues.

Focusing for conciseness on the charged sector, one can split
\begin{equation}\label{eq:chMM:split}
 \chMM=[\chMM]_2+[\chMM]_4\,,
\end{equation}
where the first term $[\chMM]_2$ contains the quadratic parameters and no vevs, while the second term $[\chMM]_4$ collects the quartic parameters and all vevs. One would need to compute the eigenvalues to diagonalize $\chMM$ and obtain the physical charged scalars. Analytically, that task is in general hopeless. There are, however, two straightforward pieces of relevant information. First, $\chMM$ has a null eigenvector $\vec c_{G}=(\vev{1},\ldots, \vev{n})^T$, corresponding to the charged would-be Goldstone boson $G^-$, since
\begin{equation}
 (\chMM \vec c_{G})_a=\partial_{\vev{a}}\vevV-\frac{i}{\vev{a}}\partial_{\vevPh{a}}\vevV=0\,.
\end{equation}
Second, the sum of squared masses of the physical charged scalars is
\begin{equation}
 \TR{\chMM}=\sum_{a=1}^n\qq{a}+[(\QQ{}\text{'s})\times(\text{vevs})^2]\,,
\end{equation}
where $[(\QQ{}\text{'s})\times(\text{vevs})^2]$ are contributions linear in $\QQ{}$'s and quadratic in vevs. For $\sum_{a=1}^n\qq{a}\gg\vev{}^2$, it follows that there are charged scalars that can be given arbitrarily large masses, that is, much larger than the electroweak scale.

In the analysis of \cite{Miro:2024zka}, in the regime where free quadratic parameters can drive large masses, one can analyze the limit $\QQ{}\to 0$, $\chMM\to[\chMM]_2$, and then treat $[\chMM]_4$ as a perturbation. In that case, one can realize that in addition to $\vec c_{G}$, another null (right) eigenvector exists, $\vec c_0=(\vev{1}e^{i2\vevPh{1}},\ldots,\vev{n}e^{i2\vevPh{n}})^T$, with 
\begin{equation}
 ([\chMM]_2\, \vec c_{0})_a=e^{i2\vevPh{a}}\left(\partial_{\vev{a}}\vevV_2+\frac{i}{\vev{a}}\partial_{\vevPh{a}}\vevV_2\right)=0\,.
\end{equation}
When quartic couplings are restored, that null eigenvalue is raised to $\mathcal O(\QQ{}\times\vev{}^2)$, thus bounded, while the would-be Goldstone remains massless. The analysis is extended to the neutral sector, accounting for 3 neutral states with bounded masses, where just one ``Higgs-like'' scalar is expected. The crux is the fact that, in addition to the vev in \cref{eq:FieldExp:01}, its CP-conjugate $\vev{a}e^{i\vevPh{a}}\mapsto \vev{a}e^{-i\vevPh{a}}$ ($a=1,\ldots,n$) is also a minimum of the potential (it is indeed as good a candidate vacuum as the chosen one): in the mass matrices, the quadratic parameters alone, in the limit of vanishing quartic parameters, cannot tell apart one minimum from the other, and there are would-be Goldstone null eigenvectors associated to both. The quartic couplings, of course, tell them apart and only the true would-be Goldstones remain null eigenvectors of the charged and neutral mass matrices.

\section{Real 3HDMs with SCPV and softly broken symmetry\label{SEC:sym3HDMSCPV}}
Following the previous discussion, real 3HDMs with SCPV provide the first class of scenarios, in terms of the number of scalar doublets involved, where the stationarity conditions leave one unconstrained quadratic parameter in the scalar potential, which might induce a (partial) decoupling regime. Attending to this fact, we consider scalar potentials
\begin{equation}\label{eq:Real3HDM:Pot:01}
 \potV(\SD{1},\SD{2},\SD{3})=\potV_2(\SD{1},\SD{2},\SD{3})+\potV_4(\SD{1},\SD{2},\SD{3})\,,
\end{equation}
where the quadratic part $\potV_2$ is the most general CP conserving possibility, as in \cref{eq:RealnHDM:V2:01},
\begin{equation}\label{eq:Real3HDM:Pot:V2:01}
 \potV_2(\SD{1},\SD{2},\SD{3})=\sum_{a=1}^3\qq{a}\SDd{a}\SD{a}+\frac{1}{2}\sum_{a=1}^{2}\sum_{b=a+1}^3\qq{ab}(\SDd{a}\SD{b}+\SDd{b}\SD{a})\,,
\end{equation}
with all 6 parameters $\qq{a}$ and $\qq{ab}$ real. For the quartic part $\potV_4$ we consider several possibilities. To maintain a reduced budget in terms of the number of parameters, we consider $\potV_4$ invariant under some discrete symmetries when the 3 doublets transform as an irreducible triplet; CP invariance is, in addition, imposed. In \cref{sSEC:sym3HDMSCPV:A4} we consider the case of the $A_4$ group: as already mentioned, with the additional CP invariance requirement, the resulting $\potV_4$ does in fact correspond to invariance under $S_4$. With a simple parametric limit, invariance under $\Delta(3n^2)$ and $\Delta(6n^2)$, with $n>3$, is obtained. In \cref{sSEC:sym3HDMSCPV:Delta27} we address some aspects of interest that arise in the case of invariance under $\Delta(27)$.

\subsection{One $A_4$ triplet with CP invariance\label{sSEC:sym3HDMSCPV:A4}}
We discuss first the scalar potential $\potV_4$ and the stationarity conditions in \cref{ssSEC:sym3HDMSCPV:A4:Pot}, and then analyze the charged and neutral scalar sectors in \cref{ssSEC:sym3HDMSCPV:A4:Charged,ssSEC:sym3HDMSCPV:A4:Neutral}, respectively.
\subsubsection{Scalar potential\label{ssSEC:sym3HDMSCPV:A4:Pot}}
The invariant quartic scalar potential is\footnote{Rather than the notation in \cref{eq:RealnHDM:V4:01}, it is more convenient to use in \cref{eq:Real3HDMA4:Pot:V4:01} quartic parameters associated to the different independent $A_4$ quartic invariants.}
\begin{equation}\label{eq:Real3HDMA4:Pot:V4:01}
 \begin{aligned}
 \potV_{4|A_4}(\SD{1},\SD{2},\SD{3})
 &= \frac{\QQ{1}}{2}\left(\sum_{a=1}^3\SDd{a}\SD{a}\right)^2+\QQ{2}\sum_{a=1}^3(\SDd{a}\SD{a})^2
   -\QQ{2}\sum_{a=1}^{2}\sum_{b=a+1}^{3}(\SDd{a}\SD{a})(\SDd{b}\SD{b})\\
 & +\QQ{3}\sum_{a=1}^2\sum_{b=a+1}^{3}(\SDd{a}\SD{b})(\SDd{b}\SD{a})
   +\frac{\QQ{4}}{2}\sum_{a=1}^{2}\sum_{b=a+1}^3\left((\SDd{a}\SD{b})^2+(\SDd{b}\SD{a})^2\right)\,,
\end{aligned}
\end{equation}
with $\QQ{j}\in\mathbb{R}$.

With the vevs in \cref{eq:FieldExp:01} and $\vev{}^2\equiv\vev{1}^2+\vev{2}^2+\vev{3}^2$, the stationarity conditions read
\begin{multline}\label{eq:Real3HDMA4:Stationarity:v:02}
 \partial_{\vev{a}}{\vevV}=\qq{a}\vev{a}+\frac{1}{2}\sum_{b< a}\qq{ba}\co{}{ba}\vev{b}+\frac{1}{2}\sum_{b>a}\qq{ab}\co{}{ab}\vev{b}\\
 +\frac{\vev{a}}{2}(\QQ{1}-\QQ{2}+\QQ{3})\vev{}^2+\frac{\vev{a}^3}{2}(3\QQ{2}-\QQ{3})+\frac{\vev{a}}{2}\QQ{4}\left(\sum_{b< a}\co{2}{ba}\vev{b}^2+\sum_{b> a}\co{2}{ab}\vev{b}^2\right)=0\,,
\end{multline}
and
\begin{equation}\label{eq:Real3HDMA4:Stationarity:th:02}
 \partial_{\vevPh{a}}{\vevV}=\frac{\vev{a}}{2}\left(\sum_{b<a}(\qq{ba}\si{}{ba}\vev{b}+\QQ{4}\si{2}{ba}\vev{a}\vev{b}^2)- \sum_{b>a}(\qq{ab}\si{}{ab}\vev{b}+\QQ{4}\si{2}{ab}\vev{a}\vev{b}^2)\right)=0\,,
\end{equation}
where, in addition to $\si{}{ab}\equiv\sin\Theta_{ab}$ and $\co{}{ab}\equiv\cos\Theta_{ab}$, with $\Theta_{ab}\equiv\vevPh{a}-\vevPh{b}$ as in \cref{eq:RealnHDM:Stationarity:v:01,eq:RealnHDM:Stationarity:th:01}, the quartic terms in \cref{eq:Real3HDMA4:Stationarity:v:02,eq:Real3HDMA4:Stationarity:th:02} involve $\si{2}{ab}\equiv\sin 2\Theta_{ab}$ and $\co{2}{ab}\equiv\cos 2\Theta_{ab}$. More explicitly, we have
\begin{align}
\label{eq:Real3HDMA4:Stationarity:v1}
 \partial_{\vev{1}}{\vevV}&=\qq{1}\vev{1}+\frac{\qq{12}}{2}c_{[12]}\vev{2}+\frac{\qq{13}}{2}c_{[13]}\vev{3}\\
 & +\frac{\vev{1}}{2}\left[\QQ{1}\vev{}^2+\QQ{2}(2\vev{1}^2-\vev{2}^2-\vev{3}^2)+\QQ{3}(\vev{2}^2+\vev{3}^2)+\QQ{4}(c_{2[12]}\vev{2}^2+c_{2[13]}\vev{3}^2)\right],\nonumber\\
\label{eq:Real3HDMA4:Stationarity:v2}
 \partial_{\vev{2}}{\vevV}&=\qq{2}\vev{2}+\frac{\qq{12}}{2}c_{[12]}\vev{1}+\frac{\qq{23}}{2}c_{[23]}\vev{3}\\
 & +\frac{\vev{2}}{2}\left[\QQ{1}\vev{}^2+\QQ{2}(2\vev{2}^2-\vev{1}^2-\vev{3}^2)+\QQ{3}(\vev{1}^2+\vev{3}^2)+\QQ{4}(c_{2[12]}\vev{1}^2+c_{2[23]}\vev{3}^2)\right],\nonumber\\
\label{eq:Real3HDMA4:Stationarity:v3}
 \partial_{\vev{3}}{\vevV}&=\qq{3}\vev{3}+\frac{\qq{13}}{2}c_{[13]}\vev{1}+\frac{\qq{23}}{2}c_{[23]}\vev{2}\\
 & +\frac{\vev{3}}{2}\left[\QQ{1}\vev{}^2+\QQ{2}(2\vev{3}^2-\vev{1}^2-\vev{2}^2)+\QQ{3}(\vev{1}^2+\vev{2}^2)+\QQ{4}(c_{2[13]}\vev{1}^2+c_{2[23]}\vev{2}^2)\right],\nonumber
\end{align}
and
\begin{align}
\label{eq:Real3HDMA4:Stationarity:th1}
 &\partial_{\vevPh{1}}{\vevV}=-\frac{\vev{1}}{2}\left(\qq{12}s_{[12]}\vev{2}+\qq{13}s_{[13]}\vev{3}\right)
  -\frac{\vev{1}^2}{2}\QQ{4}(s_{2[12]}\vev{2}^2+s_{2[13]}\vev{3}^2)\,,\\
\label{eq:Real3HDMA4:Stationarity:th2}
 &\partial_{\vevPh{2}}{\vevV}=\frac{\vev{2}}{2}\left(\qq{12}s_{[12]}\vev{1}-\qq{23}s_{[23]}\vev{3}\right)
  +\frac{\vev{2}^2}{2}\QQ{4}(s_{2[12]}\vev{1}^2-s_{2[23]}\vev{3}^2)\,,\\
\label{eq:Real3HDMA4:Stationarity:th3}
 &\partial_{\vevPh{3}}{\vevV}=\frac{\vev{3}}{2}\left(\qq{13}s_{[13]}\vev{1}+\qq{23}s_{[23]}\vev{2}\right)
  +\frac{\vev{3}^2}{2}\QQ{4}(s_{2[13]}\vev{1}^2+s_{2[23]}\vev{2}^2)\,.
\end{align}
Notice, as anticipated, that by construction $\vevV$ only depends on 2 independent vacuum phase differences rather than the 3 vacuum phases, since $\partial_{\vevPh{1}}{\vevV}+\partial_{\vevPh{2}}{\vevV}+\partial_{\vevPh{3}}{\vevV}=0$.

In terms of the quadratic parameters, one can straightforwardly read that, with \refEqs{eq:Real3HDMA4:Stationarity:th1}-\eqref{eq:Real3HDMA4:Stationarity:th3}, one can express $\qq{12}$ and $\qq{13}$ in terms of quartic parameters and $\qq{23}$. Then, with \refEqs{eq:Real3HDMA4:Stationarity:v1}-\eqref{eq:Real3HDMA4:Stationarity:v3}, one can also express $\qq{1}$, $\qq{2}$ and $\qq{3}$ in terms of quartic parameters and $\qq{23}$. That is, $\qq{23}$ is the only quadratic parameter left, in terms of which one can eventually analyze how a regime with large masses for the new scalars might be achieved. For later convenience, notice that the choice of $\qq{23}$ explicitly breaks a symmetric treatment of the 3 components of the triplet: to restore it, one can for example combine \refEqs{eq:Real3HDMA4:Stationarity:v1}-\eqref{eq:Real3HDMA4:Stationarity:v3} and choose as independent quadratic parameter
\begin{equation}
 \qq{}\equiv\qq{1}+\qq{2}+\qq{3}\,.
\end{equation}

\subsubsection{Charged scalar sector\label{ssSEC:sym3HDMSCPV:A4:Charged}}
The elements of the $3\times 3$ hermitian charged mass matrix read
\begin{equation}\label{eq:Real3HDMA4:chMM:01}
\begin{aligned}
 & (\chMM)_{aa}=\qq{a}+\frac{1}{2}(\QQ{1}-\QQ{2})\vev{}^2+\frac{3}{2}\QQ{2}\vev{a}^2\,,\\
 & (\chMM)_{ab}=(\chMM)_{ba}^\ast=\frac{e^{i\Theta_{ab}}}{2}\left[\qq{ab}+\vev{a}\vev{b}(e^{-i\Theta_{ab}}\QQ{3}+e^{i\Theta_{ab}}\QQ{4})\right]\,, \qquad a<b\,,\\
 & (\chMM)_{ab}=(\chMM)_{ba}^\ast=\frac{e^{i\Theta_{ba}}}{2}\left[\qq{ba}+\vev{a}\vev{b}(e^{-i\Theta_{ba}}\QQ{3}+e^{i\Theta_{ba}}\QQ{4})\right]\,, \qquad a>b\,.
\end{aligned}
\end{equation}
For compactness, the stationarity relations have not been used in \cref{eq:Real3HDMA4:chMM:01}. As discussed in \cref{SEC:Intro}, one can readily identify the null (left) eigenvector $\vec c_{G}$ associated to the charged would-be Goldstone boson $G^\pm$, \ie\ $\vec c_{G}=(\vev{1},\vev{2},\vev{3})^T$:
\begin{align}
 (\chMM\,\vec c_{G})_a &=\sum_{b<a}(\chMM)_{ab}\vev{b}+(\chMM)_{aa}\vev{a}+\sum_{b>a}(\chMM)_{ab}\vev{b}\nonumber\\
&=\sum_{b<a}\frac{e^{i\Theta_{ba}}}{2}\left[\qq{ba}+\vev{a}\vev{b}(e^{-i\Theta_{ba}}\QQ{3}+e^{i\Theta_{ba}}\QQ{4})\right]\vev{b} +\qq{a}\vev{a}+\frac{\vev{a}}{2}(\QQ{1}-\QQ{2})\vev{}^2+\frac{3}{2}\QQ{2}\vev{a}^3 \nonumber\\
&\,+\sum_{b>a}\frac{e^{i\Theta_{ab}}}{2}\left[\qq{ab}+\vev{a}\vev{b}(e^{-i\Theta_{ab}}\QQ{3}+e^{i\Theta_{ab}}\QQ{4})\right]\vev{b}\,,
\end{align}
that is
\begin{equation}
 (\chMM\,\vec c_{G^-})_a=\partial_{\vev{a}}{\vevV}-\frac{i}{\vev{a}}\partial_{\vevPh{a}}{\vevV}=0\,.
\end{equation}
Following \cref{eq:Real3HDMA4:chMM:01}, the sum of the squared masses of charged scalars is given by
\begin{equation}
 \TR{\chMM}=\qq{}+\frac{3}{2}\QQ{1}\vev{}^2\,,
\end{equation}
and thus $\qq{}\gg\vev{}^2$ induces (some) large masses. However, most importantly, in this case one can obtain the eigenvalues of $\chMM$ in closed form:
\begin{equation}\label{eq:Real3HDMA4:chMM:eig:01}
 \text{Eig}(\chMM)=\left\{0,\frac{1}{2}(\QQ{4}-\QQ{3})\vev{}^2,\qq{}+\frac{1}{2}(3\QQ{1}+\QQ{3}-\QQ{4})\vev{}^2\right\}\,.
\end{equation}
Clearly, \cref{eq:Real3HDMA4:chMM:eig:01} shows that, with quartic couplings bounded by perturbativity constraints, $\frac{1}{2}(\QQ{4}-\QQ{3})\vev{}^2$ cannot be much larger than the electroweak scale squared $\vev{}^2$, and this is so no matter the values of the independent quadratic parameter left, $\qq{}$. This last property manifests in a particularly transparent manner: the second eigenvalue, $\frac{1}{2}(\QQ{4}-\QQ{3})\vev{}^2$, is independent of $\qq{}$. A ``partial decoupling regime'' with $\qq{}+\frac{1}{2}(3\QQ{1}+\QQ{3}-\QQ{4})\vev{}^2\gg \frac{1}{2}(\QQ{4}-\QQ{3})\vev{}^2\sim\mathcal O(\vev{}^2)$ can be achieved through $\qq{}\gg\vev{}^2$. These features perfectly illustrate, in an analytically clean manner, the general properties discussed in \cite{Miro:2024zka} and summarized in \cref{SEC:Intro}.

In this case, tackling the determination of the charged scalar masses by separating it as in \cref{eq:chMM:split}, with quartic couplings treated as a perturbation, is unnecessary since one already has \cref{eq:Real3HDMA4:chMM:eig:01}. We nevertheless comment briefly that if one does so, the eigenvalues of $[\chMM]_2$ are simply $\{0,0,\qq{}\}$; the perturbative analysis with $[\chMM]_4$ yields the \emph{exact} result. It is worth noting that,  although it is a degenerate perturbation theory problem, in this case $[\chMM]_2$ and $[\chMM]_4$ commute once the stationarity conditions are used, $\left[[\chMM]_2,[\chMM]_4\right]=0$, and the result is particularly simple. We now turn to the neutral sector.

\subsubsection{Neutral scalar sector\label{ssSEC:sym3HDMSCPV:A4:Neutral}}
The $6\times 6$ real symmetric mass matrix in the neutral scalar sector is
\begin{equation}
 \nMM=\begin{pmatrix}M_{RR}^2 & M_{RI}^2\\ M_{IR}^2 & M_{II}^2\end{pmatrix}\,,\qquad M_{RR}^2=M_{RR}^{2\,T}\,,\ M_{II}^2=M_{II}^{2\,T}\,,\ M_{RI}^{2\,T}=M_{IR}^2\,,
\end{equation}
where the $3\times 3$ submatrices $M_{RR}^2$, $M_{II}^2$ and $M_{RI}^2$ have elements
\begin{align}
& (M_{RR}^2)_{aa}=\qq{a}+\frac{1}{2}(\QQ{1}-\QQ{2}+\QQ{3})\vev{}^2+\frac{1}{2}(2\QQ{1}+7\QQ{2}-\QQ{3})\vev{a}^2+\frac{\QQ{4}}{2}\sum_{b\neq a}\co{2}{ab}\vev{b}^2\, ,\nonumber \\
& (M_{RR}^2)_{ab}=(M_{RR}^2)_{ba}=\frac{\qq{ab}}{2}\co{}{ab}+(\QQ{1}-\QQ{2}+\QQ{3}+\QQ{4}\co{2}{ab})\vev{a}\vev{b}\,, \quad a<b\,,
\end{align}
\begin{align}
& (M_{II}^2)_{aa}=\qq{a}+\frac{1}{2}(\QQ{1}-\QQ{2}+\QQ{3})\vev{}^2+\frac{1}{2}(3\QQ{2}-\QQ{3})\vev{a}^2-\frac{\QQ{4}}{2}\sum_{b\neq a}\co{2}{ab}\vev{b}^2\, ,\nonumber \\
& (M_{II}^2)_{ab}=(M_{II}^2)_{ba}=\frac{\qq{ab}}{2}\co{}{ab}+\QQ{4}\co{2}{ab}\vev{a}\vev{b}\,, \quad a<b\,,
\end{align}
\begin{align}
& (M_{RI}^2)_{aa}=\frac{\QQ{4}}{2}\sum_{b<a}\si{2}{ba}\vev{b}^2\, -\frac{\QQ{4}}{2}\sum_{b>a}\si{2}{ab}\vev{b}^2\,,\nonumber \\
& (M_{RI}^2)_{ab}=-(M_{RI}^2)_{ba}=\frac{\qq{ab}}{2}\si{}{ab}+\QQ{4}\si{2}{ab}\vev{a}\vev{b}\,, \quad a<b\,.
\end{align}
As in the charged sector, one can readily check from the sum of squared masses of neutral scalars that
\begin{equation}
 \TR{\nMM}=2\qq{}+2(2\QQ{1}+\QQ{2}+\QQ{3})\vev{}^2\,,
\end{equation}
in such a way that, if $\qq{}\gg \vev{}^2$, there will be neutral states with masses much larger than the electroweak scale. Unfortunately, to the best of our efforts, there is no analytic solution for the roots of the characteristic polynomial $P_0(x)=\det(\nMM-x\mathbf{1})$: apart from the trivial root $x=0$ corresponding to the would-be Goldstone eigenvector $\vec n_G=(0,0,0,\vev{1},\vev{2},\vev{3})^T$, 
\begin{equation}
 (\nMM \vec n_G)_a=-\frac{1}{\vev{a}}\partial_{\vevPh{a}}\vevV=0\,,\qquad (\nMM \vec n_G)_{3+a}=\partial_{\vev{a}}\vevV=0\,,\qquad a=1,2,3,
\end{equation}
one is confronted with the calculation of the roots of a quintic polynomial. One can resort again to separate
\begin{equation}\label{eq:nMM:2:4:01}
 \nMM=[\nMM]_2+[\nMM]_4\,,
\end{equation}
where $[\nMM]_2$ and $[\nMM]_4$ collect, respectively, the contributions controlled by quadratic and quartic parameters. The eigenvalues of $[\nMM]_2$ are $\{0,0,0,0,\qq{},\qq{}\}$: the doubling of eigenvalues with respect to $[\chMM]_2$ is fully expected following \cite{Miro:2024zka}. With the perturbation $[\nMM]_4$, among the 4 null eigenvalues, 3 receive ($\QQ{}$'s)$\times \vev{}^2$ corrections. This is sufficient to prove boundedness following perturbativity constraints on $\QQ{}$'s, but provides no further insight beyond the general argument of \cite{Miro:2024zka}. There is, however, a kind-of intermediate regime in this $A_4$ invariant scenario. One can notice that the commutator of $[\nMM]_2$ and $[\nMM]_4$, using the stationarity conditions, is
\begin{equation}
\left[[\nMM]_2,[\nMM]_4\right]\propto(3\QQ{2}-\QQ{3}-\QQ{4})\,(\ldots)\,,
\end{equation}
where the right-hand side is a global factor $(3\QQ{2}-\QQ{3}-\QQ{4})$ times a $6\times 6$ matrix. It follows that if one splits, for example, $\QQ{3}=3\QQ{2}-\QQ{4}+\delta_3$, it is now possible to separate
\begin{equation}
 \nMM=[\nMM]_{\delta_3=0}+[\nMM]_{\delta_3}\,,
\end{equation}
where $[\nMM]_{\delta_3=0}$ would be the full, exact mass matrix if $\QQ{3}=3\QQ{2}-\QQ{4}$, 
and treat $[\nMM]_{\delta_3}$ as a perturbation of $\nMM=[\nMM]_{\delta_3=0}$, which already incorporates quartic parameter contributions that can lift the degeneracies and masslessness of $[\nMM]_2$ eigenstates. Despite corresponding again to the obtention of roots of a quintic polynomial, contrary to the case of $\nMM$, one can obtain the eigenvalues of $[\nMM]_{\delta_3=0}$, which are
\begin{align}
&M_{G^0}=0\,,\\
&M_{\Fnf{1}}^2=\QQ{4}\vev{}^2\,,\label{eq:eig:nMMspec:01}\\
&M_{\Fnf{2,3}}^2=\frac{1}{2}(\QQ{1}+2\QQ{2})\vev{}^2\pm\frac{1}{2}\sqrt{(\QQ{1}+2\QQ{2})^2\vev{}^4+16\QQ{4}(\QQ{4}-\QQ{1}-2\QQ{2})\Sigma}\,,\label{eq:eig:nMMspec:02}\\
&M_{\Fnf{4,5}}^2=\qq{}+\frac{3}{2}(\QQ{1}+2\QQ{2}-\QQ{4})\vev{}^2\pm\frac{\QQ{4}}{2}\sqrt{\vev{}^4-4\Sigma}\,,\label{eq:eig:nMMspec:03}
\end{align}
where $\Sigma=\si{}{12}^2\vev{1}^2\vev{2}^2+\si{}{13}^2\vev{1}^2\vev{3}^2+\si{}{23}^2\vev{2}^2\vev{3}^2$. As expected, there are 3 eigenvalues in \cref{eq:eig:nMMspec:01,eq:eig:nMMspec:02} that remain bounded even if $\qq{}\gg\vev{}^2$ (they are indeed independent of $\qq{}$), while 2 eigenvalues, in \refEq{eq:eig:nMMspec:03}, can be made arbitrarily large in that (partial) decoupling regime. 
One can now introduce $[\nMM]_{\delta_3}$ as a perturbation; it is to be noticed that the situation is simpler than in the discussion after \cref{eq:nMM:2:4:01}, since it is no longer a degenerate perturbation theory problem.

\subsubsection{The $\Delta(3n^2)$ and $\Delta(6n^2)$, with $n>3$, limit\label{sSEC:sym3HDMSCPV:S4:Delta}}
The potential corresponding to invariance under $\Delta(3n^2)$, $\Delta(6n^2)$, with $n>3$, is simply obtained setting $\QQ{4}\to 0$ in \cref{eq:Real3HDMA4:Pot:V4:01}. This change propagates straightforwardly to the charged scalar mass spectrum in \cref{eq:Real3HDMA4:chMM:eig:01}. Concerning the neutral sector, the mass spectrum is again beyond analytic reach; the intermediate regime leading to \refEqs{eq:eig:nMMspec:01}-\eqref{eq:eig:nMMspec:03} in the $A_4$ scenario is much less interesting for $\QQ{4}\to 0$: one can notice that \cref{eq:eig:nMMspec:01,eq:eig:nMMspec:02} only give one non-vanishing eigenvalue (rather than all three), and the eigenvalues in \cref{eq:eig:nMMspec:03} become degenerate.

\subsection{One $\Delta(27)$ triplet with CP invariance\label{sSEC:sym3HDMSCPV:Delta27}}
The $\Delta(27)$ invariant quartic potential for an irreducible triplet is
\begin{align}\label{eq:Real3HDMD27:Pot:V4:01}
 \potV_{4|\Delta(27)}(\SD{1},\SD{2},\SD{3})
 &= \frac{\QQ{1}}{2}\sum_{a=1}^3\left(\SDd{a}\SD{a}\right)^2+\QQ{2}\sum_{a,b=1}^{3}(\SDd{a}\SD{a})(\SDd{b}\SD{b})+\QQ{3}\sum_{a,b=1}^{3}(\SDd{a}\SD{b})(\SDd{b}\SD{a}) \nonumber \\
 &  +\frac{\QQ{\Delta}}{2}\left[(\SDd{1}\SD{2})(\SDd{1}\SD{3})+(\SDd{2}\SD{3})(\SDd{2}\SD{1})+(\SDd{3}\SD{1})(\SDd{3}\SD{2})\right] \nonumber \\
 &  +\frac{\QQ{\Delta}^\ast}{2}\left[(\SDd{2}\SD{1})(\SDd{3}\SD{1})+(\SDd{3}\SD{2})(\SDd{1}\SD{2})+(\SDd{1}\SD{3})(\SDd{2}\SD{3})\right],
\end{align}
and imposing CP invariance one has $\QQ{\Delta}^\ast=\QQ{\Delta}$.

In the $\Delta(27)$ case, concerning analytic results, we exclusively focus on the charged sector since this scenario is not as transparent as the $A_4$ case in \cref{sSEC:sym3HDMSCPV:A4}, but still of interest. The characteristic polynomial of $\chMM$ reads
\begin{equation}
 P_{\pm}(x)=\det(\chMM-x\mathbf{1})=-x(x^2+c_2x+c_1)\,.
\end{equation}
It is straightforward to obtain the coefficient $c_2$ as
\begin{equation}
 -c_2=\TR{\chMM}=\qq{}+\frac{1}{2}(\QQ{1}+6\QQ{2}+2\QQ{3})\vev{}^2\,,
\end{equation}
while the expression for $c_1$ is quite cumbersome. The point is that it has the following form
\begin{equation}
 c_1= \qq{}\left[\QQ{3}\vev{}^2+\QQ{\Delta}\mathcal O(\vev{}^2)\right]+\mathcal O(\QQ{}^2\vev{}^4)\,,
\end{equation}
where $\mathcal O(\vev{}^2)$ is a quadratic expression in the vevs $\vev{a}$, while $\mathcal O(\QQ{}^2\vev{}^4)$ is quadratic in the quartic parameters $\QQ{a}$ and quartic in the vevs $\vev{a}$.

The squared masses of the charged scalars are the non-null roots of $P_\pm(x)$,
\begin{equation}
 \frac{1}{2}\left(-c_2\pm\sqrt{c_2^2-4c_1}\right)\,,
\end{equation}
where now
\begin{equation}
 c_2^2-4c_1=(\qq{})^2+\qq{}\left[(\QQ{1}+6\QQ{2}-2\QQ{3})\vev{}^2-4\QQ{\Delta}\mathcal O(\vev{}^2)\right]-4\mathcal O(\QQ{}^2\vev{}^4)\,.
\end{equation}
Contrary to the $A_4$ case where the bounded charged scalar mass in \cref{eq:Real3HDMA4:chMM:eig:01} is transparently independent of $\qq{}$, in the $\Delta(27)$ case \emph{it is not}, but it is nevertheless clear that in the regime $\qq{}\gg\vev{}^2$, the two squared mass eigenvalues become
\begin{equation}\label{eq:Real3HDMDelta27:chMM:eig:01}
 M_{\FchfPM{1}}^2\;\to\; \QQ{3}\vev{}^2+\mathcal O(\QQ{\Delta}\vev{}^2),\,\qquad M_{\FchfPM{2}}^2\;\to\; \qq{}+\frac{1}{2}(\QQ{1}+6\QQ{2})\vev{}^2-\mathcal O(\QQ{\Delta}\vev{}^2)\,.
\end{equation}
The interesting aspect is that \cref{eq:Real3HDMDelta27:chMM:eig:01} illustrates again the boundedness of the masses of part of the new scalars, in a scenario in which they are under analytic control, although not independent of the free quadratic parameter left.

\section{Numerical exploration and phenomenology\label{SEC:NumPheno}}
In the previous section, we have explored in detail how part of the mass spectrum is necessarily bounded in some illustrative 3HDM scenarios with a CP invariant potential --- including a symmetry-shaped quartic part ---, and spontaneous CP violation. Although some aspects are analytically affordable, \eg\ the charged scalar masses in \cref{ssSEC:sym3HDMSCPV:A4:Charged}, a more complete analysis of the parameter space (or rather the space of these classes of models) requires a numerical approach which should incorporate some relevant constraints and cover some phenomenological consequences of potential interest. Further details of the numerical procedure are described in App.~\ref{APP:PotNum}.

Our focus is on properties that derive from the scalar sector alone, that is, from (i) the scalar potential, and (ii) the gauge interactions of the scalars. The masses and mixings of the physical scalars are obtained, as discussed extensively, from the scalar potential (and spontaneous symmetry breaking), together with cubic and quartic scalar interactions. The gauge interactions are incorporated as usual through the kinetic terms with covariant derivatives. 
Yukawa couplings of the fermions with $\SD{a}$'s are certainly necessary to provide fermion masses (and a host of Yukawa interactions with charged and neutral physical scalars), but they are beyond the scope of this work, since related constraints are not straightforwardly related to the scalar sector alone. 
Furthermore, we assume for simplicity that the lightest neutral scalar is the Higgs-like state with mass $\sim 125$ GeV \cite{ATLAS:2012yve,CMS:2012qbp}, labelled $h$ in the following. This is just a convenient assumption. 
The mass spectra of the charged and neutral scalars are, of course, of immediate interest. Beyond the mass spectra, the focus should be on gauge interactions that involve one or two scalars, specially the ones with bounded masses, providing both constraints and observables of interest. The term $(D_\mu\SD{a})^\dagger(D^\mu\SD{a})$ with $D_\mu$ the covariant derivative, includes\footnote{There is also a term $\frac{g\vev{a}}{2}\left[\frac{g}{2c_W}(1-2s_W^2)Z^\mu+eA^\mu\right]\left(W_\mu^-\chfP{a}+W_\mu^+\chfM{a}\right)$, but on summation over $a$, one is left with the would-be Goldstone contribution alone, $\vev{a}\to\vev{}$ and $\chfPM{a}\to G^\pm$.}
\begin{align}\label{eqn:covariant-derivative}
(D_\mu\SD{a})^\dagger(D^\mu\SD{a}) &\supset \left(\frac{g^2\vev{a}}{2}W_\mu^+W^{-\mu}+\frac{g^2\vev{a}}{4c_W^2}Z_\mu Z^\mu\right)\,\nfR{a} \nonumber \\
&+ i\frac{g}{2}\left(W_\mu^-\partial^\mu\chfP{a}-W_\mu^+\partial^\mu\chfM{a}\right)\nfR{a}+i\frac{g}{2}\left(W_\mu^+\chfM{a}-W_\mu^-\chfP{a}\right)\partial^\mu\nfR{a} \nonumber \\
&+\frac{g}{2}\left(W_\mu^-\partial^\mu\chfP{a}+W_\mu^+\partial^\mu\chfM{a}\right)\nfI{a}-\frac{g}{2}\left(W_\mu^+\chfM{a}+W_\mu^-\chfP{a}\right)\partial^\mu\nfI{a} \nonumber \\
&+\frac{g}{2c_W}Z^\mu\left(\nfI{a}\partial_\mu\nfR{a}-\nfR{a}\partial_\mu\nfI{a}\right)\,.
\end{align}
The physical scalars --- charged $\FchfPM{1}$, $\FchfPM{2}$, and neutral $h$, $\Fnf{j}$, $j=1,\ldots,4$ --- are obtained through the diagonalization of the mass terms in \refEqs{eq:pot:massterms:01}-\eqref{eq:nMM}. On the one hand, in the charged sector we have
\begin{equation}\label{eq:MdiagCh:01}
\UmatDag\,\chMM\,\Umat=\text{Diag}(0,M^2_{\FchfPM{1}},M^2_{\FchfPM{2}})\,,\quad M_{\FchfPM{1}}\leq M_{\FchfPM{2}}\,,
\end{equation}
where $U$ is the $3\times 3$ unitary matrix relating $\chfvec=\Umat\,\CHfvec$, with $\chfvec^T=(\chfM{1},\chfM{2},\chfM{3})$ and $\CHfvec^T=(G^-,\FchfM{1},\FchfM{2})$. On the other hand, in the neutral sector one has
\begin{equation}\label{eq:MdiagN:01}
\OmatT\,\nMM\,\Omat=\text{Diag}(0,m_h^2,M^2_{\Fnf{1}},M^2_{\Fnf{2}},M^2_{\Fnf{3}},M^2_{\Fnf{4}})\,,\quad m_h^2\leq M^2_{\Fnf{1}}\,,\quad M^2_{\Fnf{j}}\leq M^2_{\Fnf{k}}\ \ \text{for}\ \ j<k\,,
\end{equation}
where $O$ is the $6\times 6$ real orthogonal that relates $\nfvec=\Omat\,\Nfvec$, with $\nfvec^T=(\nfR{1},\nfR{2},\nfR{3},\nfI{1},\nfI{2},\nfI{3})$ and $\Nfvec^T=(G^0,h,\Fnf{1},\Fnf{2},\Fnf{3},\Fnf{4})$.

From the complete kinetic term $\sum_{a=1}^3(D_\mu\SD{a})^\dagger(D^\mu\SD{a})$ and the first line in \cref{eqn:covariant-derivative}, the interaction of the Higgs-like $h$ and vector boson pairs reads
\begin{equation}\label{eq:VVN:02}
 \left(\frac{1}{\vev{}}\sum_{a=1}^3\vev{a}\Omatel{a2}\right)\left(\frac{g^2\vev{}}{2}W_\mu^+W^{-\mu}+\frac{g^2\vev{}}{4c_W^2}Z_\mu Z^\mu\right)\,h\,.
\end{equation}
The $hW^+W^-$ and $hZZ$ coupling relative to the SM value, $K_{hVV}$, is simply obtained from the first factor,
\begin{equation}\label{eq:hVV:01}
K_{hVV}=\frac{1}{\vev{}}\sum_{a=1}^n\vev{a}\Omatel{a2}\,.
\end{equation}
Similarly, we introduce
\begin{equation}\label{eq:HVV:01}
 K_{\Fnf{1}VV}=\frac{1}{\vev{}}\sum_{a=1}^n\vev{a}\Omatel{a3}\,,\qquad 
 K_{\Fnf{2}VV}=\frac{1}{\vev{}}\sum_{a=1}^n\vev{a}\Omatel{a4}\,.
\end{equation}
$K_{hVV}$ is highly constrained from Higgs-related analyses \cite{CMS:2018uag,ATLAS:2019nkf,ParticleDataGroup:2024cfk}. As a constraint in our numerical exploration we consider
\begin{equation}
K_{hVV}\geq 0.90\,.
\end{equation}
A second constraint stems from electroweak precision data analyses in terms of the oblique parameters \cite{Altarelli:1990zd,Peskin:1991sw}. According to \cite{ParticleDataGroup:2024cfk}, we consider the following deviations with respect to the SM values
\begin{equation}\label{eq:STU:val}
 \left.\Delta S\right|_{\rm ref}=0.021\pm 0.096\,,\quad \left.\Delta T\right|_{\rm ref}=0.04\pm 0.12\,,\quad \left.\Delta U\right|_{\rm ref}=0.008\pm 0.092\,,
\end{equation}
\begin{equation}\label{eq:STU:cor}
 \rho_{ST}=0.91\,,\quad \rho_{SU}=-0.62\,,\quad \rho_{TU}=-0.83\,,
\end{equation}
where central values with corresponding uncertainties are given in \cref{eq:STU:val} and their correlations in \cref{eq:STU:cor}. We use the predictions for $\Delta S$, $\Delta T$ and $\Delta U$ computed as in \cite{Grimus:2008nb}, and impose $\chi^2(\Delta S,\Delta T,\Delta U)\leq 6.251$, where
\begin{equation}
 \chi^2(\Delta S,\Delta T,\Delta U)=\vec\delta^{\,T}\,(\mathrm{Cov}^{-1})\,\vec\delta\,,\;\; \text{ with }\;\;\vec\delta^{\,T}=(\Delta S-\left.\Delta S\right|_{\rm ref},\Delta T-\left.\Delta T\right|_{\rm ref},\Delta U-\left.\Delta U\right|_{\rm ref})\,,
\end{equation}
and $\mathrm{Cov}$ is the covariance matrix. The choice $\chi^2\leq 6.251$ corresponds to a 90 \% CL in a 3-dimensional $\chi^2$ distribution (in the 1-dimensional case that is the interval up to a $\sim 1.6\sigma$ deviation).

One last constraint concerns the mass of the lightest charged scalar $M_{\FchfPM{1}}$. Both $M_{\FchfPM{1}}$ and the Yukawa couplings of $\FchfPM{1}$ are typically very constrained by flavor changing processes. Without specifying the Yukawa couplings and thus not imposing flavor constraints, one might have light $\FchfPM{1}$ unimpeded, with \eg\ $M_{\FchfPM{1}}<M_W$. There is, however, a process that is relevant irrespective of how the Yukawa sector is handled. From the quartic terms in the scalar potential, a coupling $h\FchfP{1}\FchfM{1}$ arises, and $h\to\FchfP{1}\FchfM{1}$ is kinematically allowed if $M_{\FchfPM{1}}<m_h/2$: the corresponding width $\Gamma(h\to\FchfP{1}\FchfM{1})$ would be quite constrained. One can impose, for simplicity, $M_{\FchfPM{1}}>m_h/2$. There are, however, more restrictive and rather model-independent constraints obtained at LEP2 \cite{ALEPH:2013htx}: we incorporate them imposing $M_{\FchfPM{1}}>90$ GeV.

In the following, we discuss different results, focusing on the $A_4$-shaped scenario of \cref{sSEC:sym3HDMSCPV:A4}. Prior to the full numerical exploration, which includes the constraints on $K_{hVV}$ and on the oblique parameters, it is worth checking that, removing those constraints:
\begin{itemize}
\item $K_{hVV}$ can span the whole available range $[-1;1]$, and this is so no matter if the heaviest scalars have masses much larger than 1 TeV or not. This confirms the expectation that this scenario is not naturally aligned, and $K_{hVV}$ provides a relevant constraint.
\item $\chi^2(\Delta S,\Delta T,\Delta U)$ can be (much) larger than the imposed bound, except for the regime with all scalar masses below 1 TeV, where the bound appears to be easily satisfied; it will be relevant in the partial decoupling regime.
\end{itemize}

\begin{figure}[!ht]
\centering
\subfloat[$M_{\Fnf{2}}$ vs. $M_{\Fnf{1}}$.\label{sfig:Masses:M02vsM01}]{\includegraphics[width=0.45\textwidth]{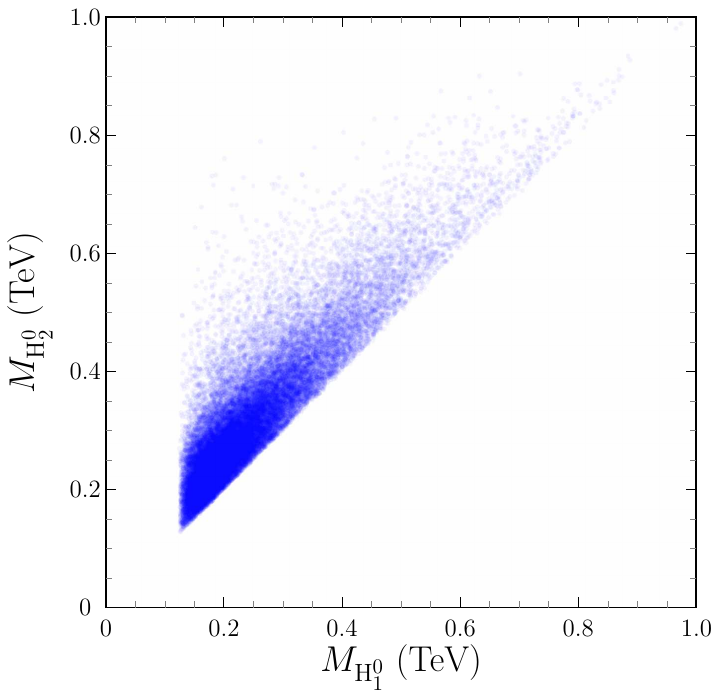}}\qquad 
\subfloat[$M_{\Fnf{4}}$ vs. $M_{\Fnf{3}}$.\label{sfig:Masses:M04vsM03}]{\includegraphics[width=0.45\textwidth]{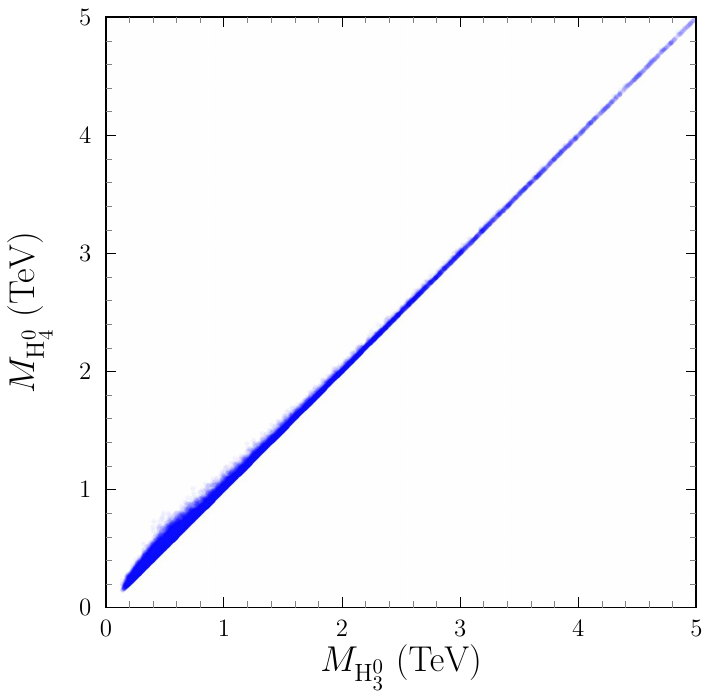}}\\
\subfloat[$M_{\FchfPM{1}}$ vs. $M_{\Fnf{1}}$.\label{sfig:Masses:MC1vsM01}]{\includegraphics[width=0.45\textwidth]{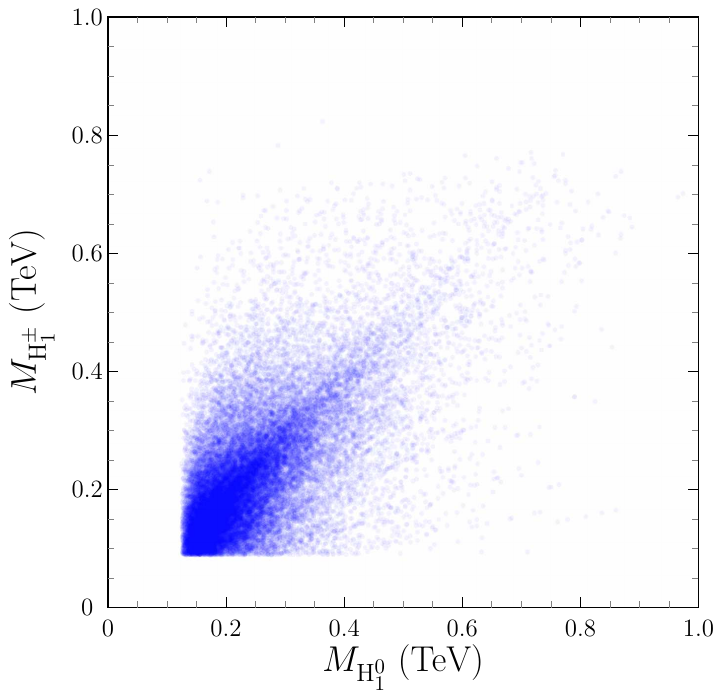}}\qquad 
\subfloat[$M_{\FchfPM{2}}$ vs. $M_{\Fnf{4}}$.\label{sfig:Masses:MC2vsM04}]{\includegraphics[width=0.45\textwidth]{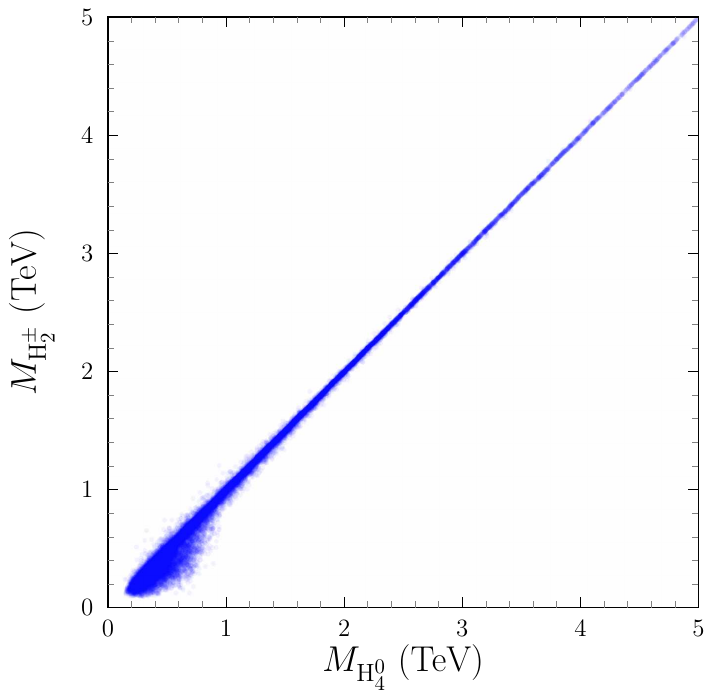}}
\caption{Masses of the new scalars.\label{fig:Masses}} 
\end{figure}

Moving into the results of the full numerical exploration, \cref{fig:Masses} shows scatter plots\footnote{We opt for scatter plots because they directly illustrate the allowed regions without further processing of the numerical output; it is important to stress that we do not attach any statistical meaning or interpretation to the density of points, which derives from the sampling procedure.} of different pairs of masses. \cref{sfig:Masses:M02vsM01} clearly shows that, as expected, both $M_{\Fnf{1}}$ and $M_{\Fnf{2}}$ are bounded, at the level of $\sim 800$ GeV. On the contrary, as \cref{sfig:Masses:M04vsM03} illustrates, $M_{\Fnf{3}}$ and $M_{\Fnf{4}}$ might be much heavier, and when that is the case they are degenerate. Concerning the charged scalars, in \cref{sfig:Masses:MC1vsM01} one can check that $M_{\FchfPM{1}}$ is bounded as expected, $M_{\FchfPM{1}}<800$ GeV, and that the constraint $M_{\FchfPM{1}}>90$ GeV is effectively at work. On the contrary, according to \cref{sfig:Masses:MC1vsM01}, $\FchfPM{2}$ can be arbitrarily heavy, and in that regime it is degenerate with $\Fnf{3}$ and $\Fnf{4}$. One last fact concerning \cref{fig:Masses} is that a regime with all new scalars as light as we allow within the numerical exploration (neutral scalars heavier than $m_h$, charged scalars heavier than $90$ GeV) appears to be perfectly feasible from considerations on the scalar sector alone.

\begin{figure}[!ht]
\centering
\subfloat[$|K_{\Fnf{1}VV}|^2$ vs. $M_{\Fnf{1}}$.\label{sfig:SVV:H1VVvsM01}]{\includegraphics[width=0.3\textwidth]{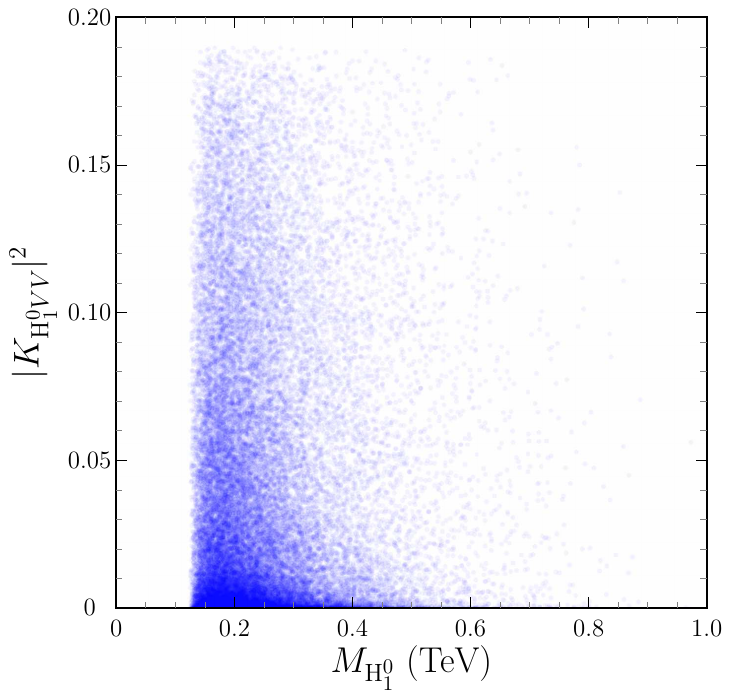}}\qquad 
\subfloat[$|K_{\Fnf{2}VV}|^2$ vs. $M_{\Fnf{2}}$.\label{sfig:SVV:H2VVvsM02}]{\includegraphics[width=0.3\textwidth]{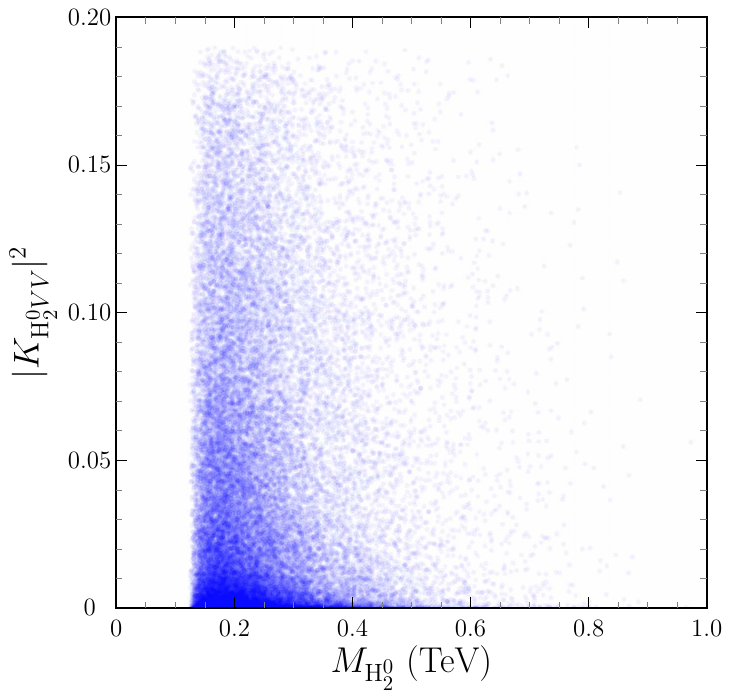}}\qquad
\subfloat[$|K_{\Fnf{2}VV}|^2$ vs. $|K_{\Fnf{1}VV}|^2$.\label{sfig:SVV:H2VVvsH1VV}]{\includegraphics[width=0.3\textwidth]{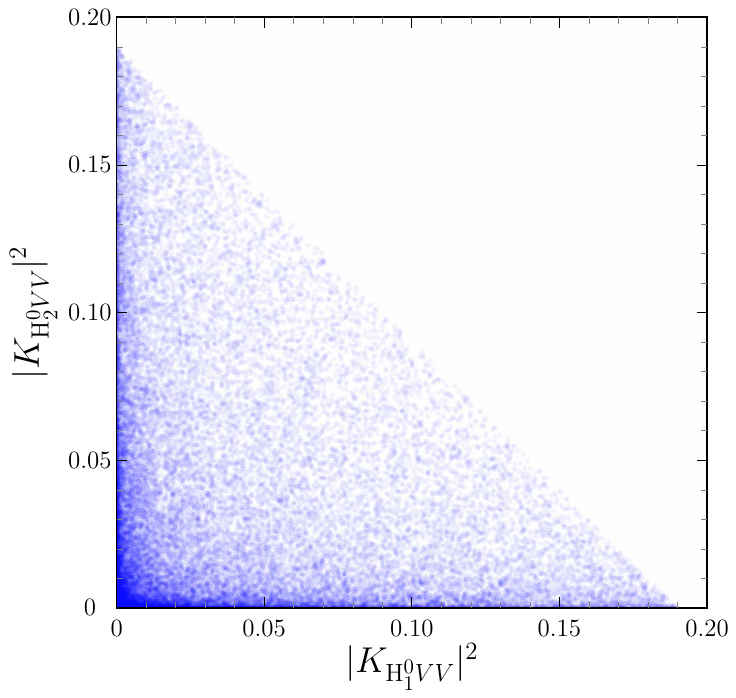}}
\caption{$\Fnf{1,2}$ couplings $W^+W^-$, $ZZ$.\label{fig:SVV}} 
\end{figure}

The next quantities of interest are $K_{\Fnf{1}VV}$ and $K_{\Fnf{2}VV}$, relevant both for production and decays $\Fnf{1},\Fnf{2}\to W^+W^-,ZZ$. The point is that the corresponding decay widths (when the processes are kinematically allowed) have a straightforward dependence on the decaying particle mass, and thus one can focus on the couplings themselves, in particular (i) what are their allowed ranges, (ii) if these allowed ranges exhibit some dependence on the decaying particle mass, and (iii) what is the interplay/correlation among the different couplings. \cref{sfig:SVV:H1VVvsM01,sfig:SVV:H2VVvsM02} show the allowed range $[0;0.19]$ which, importantly, does not appear to depend on the mass. Orthogonality of the mixing matrix $\Omat$ imposes the upper limit $|K_{\Fnf{1}VV}|^2+|K_{\Fnf{2}VV}|^2\leq 1 - |K_{hVV}|^2$: with the requirement $K_{hVV}\geq 0.90$, the upper limit value $0.19$ follows, together with the joint allowed region in \cref{sfig:SVV:H2VVvsH1VV}. As a final comment concerning \cref{sfig:SVV:H2VVvsH1VV}, it is clear that it is perfectly possible that either $K_{\Fnf{1}VV}$, or $K_{\Fnf{2}VV}$, or both, can be quite suppressed.

In terms of potential interest, the next set of observables or quantities to be addressed are the couplings that control processes such as $\Fnf{1}\to Zh$, $\Fnf{2}\to Zh$, $\FchfPM{1}\to W^\pm h$. With the interaction terms in \cref{eqn:covariant-derivative} and the mixing matrices in \refEqs{eq:MdiagCh:01}-\eqref{eq:MdiagN:01}, the decay widths of $S_i\to VS_f$, with $S_i$, $S_f$ scalars, and a (massive) vector boson $V$, are
\begin{equation}
 \Gamma(S_i\to VS_f)=|\widehat{C}_{S_1VS_2}|^2\frac{M_{S_i}^3}{64\pi\,M_V^2}\left[1-\left(\frac{M_{S_f}+M_V}{M_{S_i}}\right)^2\right]^{\frac{3}{2}}\left[1-\left(\frac{M_{S_f}-M_V}{M_{S_i}}\right)^2\right]^{\frac{3}{2}}\,,
\end{equation}
where 
\begin{equation}\label{eq:H1Zh}
 \widehat{C}_{\Fnf{1}Zh}=\frac{g}{2c_W}C_{\Fnf{1}Zh}\,,\quad C_{\Fnf{1}Zh}=\sum_{a=1}^3\left(\Omatel{a2}\Omatel{3+a,3}-\Omatel{a3}\Omatel{3+a,2}\right)\,,
\end{equation}
and similarly
\begin{equation}\label{eq:H2Zh:H2ZH1}
 C_{\Fnf{2}Zh}=\sum_{a=1}^3\left(\Omatel{a2}\Omatel{3+a,4}-\Omatel{a4}\Omatel{3+a,2}\right)\,,\qquad C_{\Fnf{2}Z\Fnf{1}}=\sum_{a=1}^3\left(\Omatel{a3}\Omatel{3+a,4}-\Omatel{a4}\Omatel{3+a,3}\right)\,,
\end{equation}
\begin{align}\label{eq:CH1Wh}
 &\widehat{C}_{\FchfPM{1}W^\mp h}=g\,C_{\FchfPM{1}W^\mp h}\,,\qquad \text{with}\qquad C_{\FchfPM{1}W^\mp h}=\sum_{a=1}^3\Umatcel{a2}\left(i\Omatel{a2}+\Omatel{3+a,2}\right)\,,\nonumber\\
  &C_{\FchfPM{1}W^\mp \Fnf{1}}=\sum_{a=1}^3\Umatcel{a2}\left(i\Omatel{a3}+\Omatel{3+a,3}\right)\,,\qquad
 C_{\FchfPM{1}W^\mp \Fnf{2}}=\sum_{a=1}^3\Umatcel{a2}\left(i\Omatel{a4}+\Omatel{3+a,4}\right)\,.
\end{align}
As in the case of $K_{\Fnf{1}VV}$, $K_{\Fnf{2}VV}$, the dependence on the masses of the corresponding decay widths is straightforward and one can focus on the couplings in \refEqs{eq:H1Zh}-\eqref{eq:CH1Wh} themselves, with allowed ranges similarly constrained by orthogonality of the mixing matrix $\Omat$ --- we do not show, for instance, $C_{\Fnf{1}Zh}$ vs. $M_{\Fnf{1}}$, $C_{\Fnf{2}Zh}$ vs. $M_{\Fnf{2}}$, and $C_{\Fnf{2}Zh}$ vs. $C_{\Fnf{1}Zh}$ plots since they are hardly distinguishable from \refFigs{sfig:SVV:H1VVvsM01}-\ref{sfig:SVV:H2VVvsH1VV}. In particular, (i) $C_{\Fnf{1}Zh}$ and $C_{\Fnf{2}Zh}$ cannot be simultaneously large (within the allowed range), and (ii) nothing prevents both being small simultaneously.

\begin{figure}[!ht]
\centering
\subfloat[$|C_{\Fnf{2}Zh}|^2$ vs. $|K_{\Fnf{1}VV}|^2$.\label{sfig:H2ZhvsH1VV:NoDec}]{\includegraphics[width=0.45\textwidth]{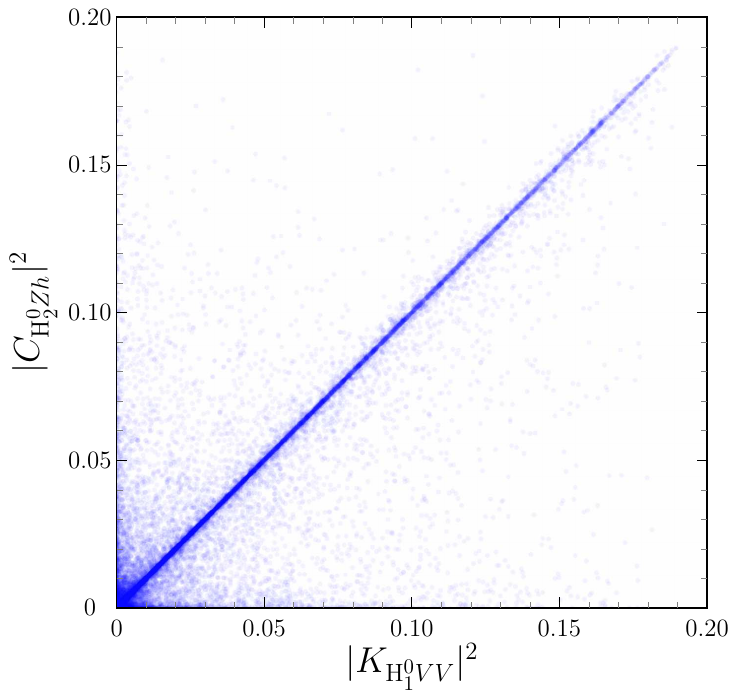}}\qquad 
\subfloat[$|C_{\Fnf{2}Z\Fnf{1}}|^2$ vs. $|C_{\Fnf{1}Zh}|^2$.\label{sfig:H2ZH1vsH1Zh:NoDec}]{\includegraphics[width=0.45\textwidth]{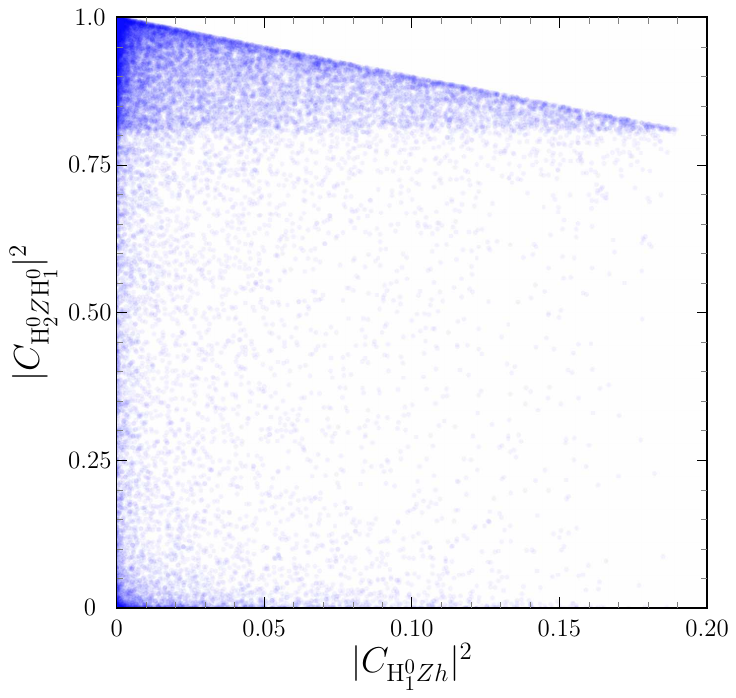}}
\caption{Interplay among couplings, $M_{\FchfPM{2}}<1$ TeV. \label{fig:Mixed1:NoDec}} 
\end{figure}

\begin{figure}[!ht]
\centering
\subfloat[$|C_{\Fnf{2}Zh}|^2$ vs. $|K_{\Fnf{1}VV}|^2$.\label{sfig:H2ZhvsH1VV:Dec}]{\includegraphics[width=0.45\textwidth]{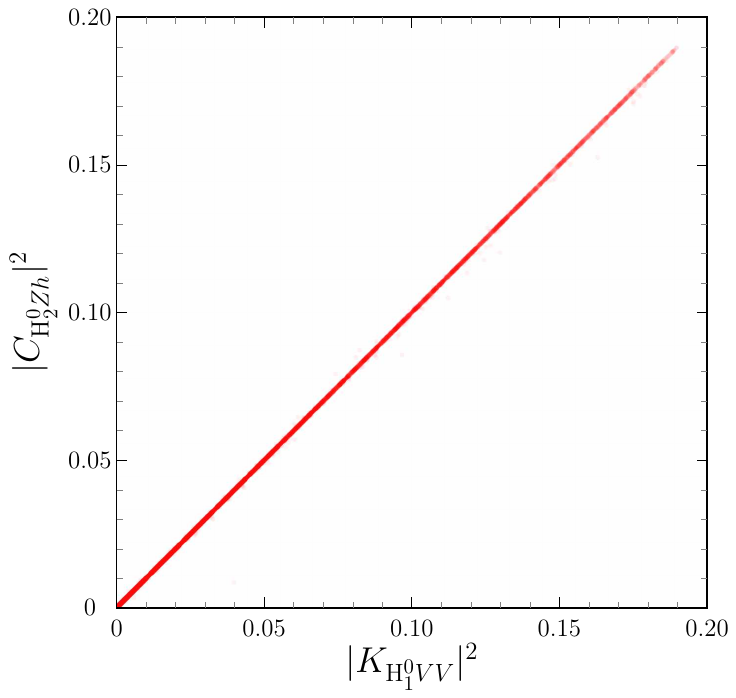}}\qquad 
\subfloat[$|C_{\Fnf{2}Z\Fnf{1}}|^2$ vs. $|C_{\Fnf{1}Zh}|^2$.\label{sfig:H2ZH1vsH1Zh:Dec}]{\includegraphics[width=0.45\textwidth]{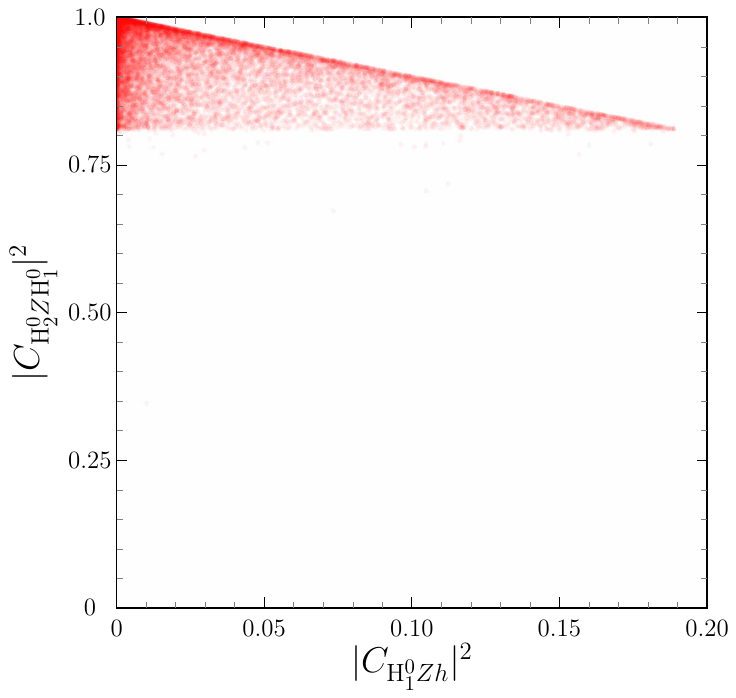}}
\caption{Interplay among couplings, $M_{\FchfPM{2}}>1$ TeV. \label{fig:Mixed1:Dec}} 
\end{figure}

The previous discussion gives a rather simple separate picture of the couplings $K_{\Fnf{1,2}VV}$ on one hand, and $C_{\Fnf{1,2}Zh}$ on the other hand. The interplay among different couplings has more interesting aspects, as illustrated in \cref{fig:Mixed1:NoDec,fig:Mixed1:Dec}. \refFig{fig:Mixed1:NoDec} shows the joint available regions of $|C_{\Fnf{2}Zh}|^2$ vs. $|K_{\Fnf{1}VV}|^2$ and of $|C_{\Fnf{2}Z\Fnf{1}}|^2$ vs. $|C_{\Fnf{1}Zh}|^2$, for $M_{\FchfPM{2}}<1$ TeV, that is, when the whole scalar spectrum is ``light''. One can observe that the whole orthogonality-of-$\Omat$-allowed region is populated. Furthermore, it is clear that there is some pattern/concentration arising in the parameter sampling that can be understood from \cref{fig:Mixed1:Dec}, which shows the same plots for $M_{\FchfPM{2}}>1$ TeV, that is, when $\FchfPM{2}$, $\Fnf{3}$ and $\Fnf{4}$ are in the decoupling regime. In this case, rather than the whole region observed for $M_{\FchfPM{2}}<1$ TeV, it is remarkable that in \cref{sfig:H2ZhvsH1VV:Dec}, $|C_{\Fnf{2}Zh}|^2\simeq|K_{\Fnf{1}VV}|^2$ (similarly, although not shown, $|C_{\Fnf{1}Zh}|^2\simeq|K_{\Fnf{2}VV}|^2$). The interesting feature of \cref{sfig:H2ZH1vsH1Zh:Dec} is that $|C_{\Fnf{2}Z\Fnf{1}}|^2$ must be large in that regime. The key is that, having fixed the ordering of the mass eigenvalues, the regime with $M_{\FchfPM{2}},M_{\Fnf{3}},M_{\Fnf{4}}>1$ TeV appears to give a rather simple correspondence between masses and mixing matrices, since they ultimately depend on the same parameters in $\nMM$ and $\chMM$. If one then lowers the values of the heavy masses, nothing appears to prevent that this picture is messed up through the eigenvalue ordering. This can be better understood in the charged sector: while it is clear that for $M_{\FchfPM{2}}>1$ TeV, $M_{\FchfPM{2}}$ corresponds to the $\qq{}$-dependent eigenvalue in \cref{eq:Real3HDMA4:chMM:eig:01}, this might not be the case for $M_{\FchfPM{2}}<1$ TeV.

\begin{figure}[!ht]
\centering
\subfloat[$|C_{\FchfPM{1}W^\mp\Fnf{1}}|^2$ vs. $M_{\FchfPM{1}}$.\label{sfig:CH1WH1vsMC1:NoDec}]{\includegraphics[width=0.31\textwidth]{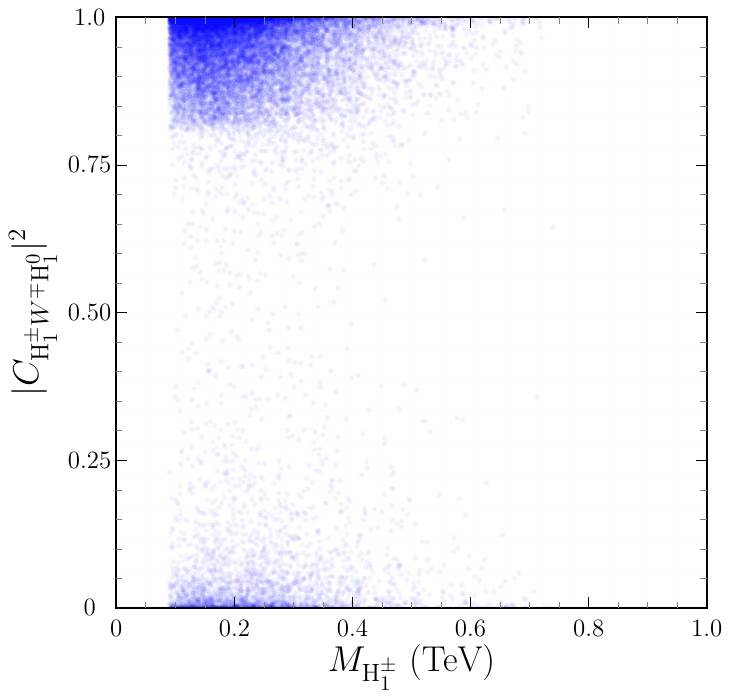}}\quad
\subfloat[$|C_{\FchfPM{1}W^\mp\Fnf{1}}|^2$ vs. $|C_{\FchfPM{1}W^\mp h}|^2$.\label{sfig:CH1WH1vsCH1Wh:NoDec}]{\includegraphics[width=0.31\textwidth]{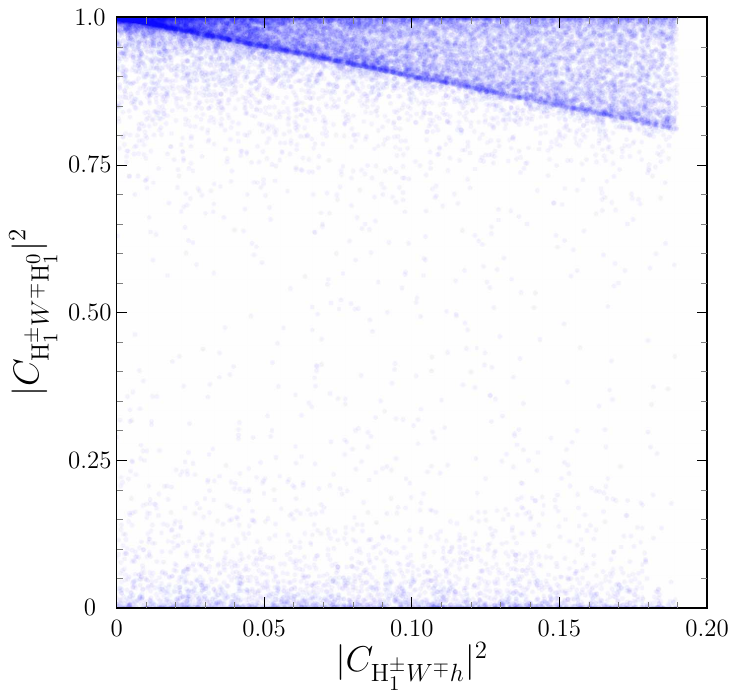}}\quad
\subfloat[$|C_{\FchfPM{1}W^\mp\Fnf{2}}|^2$ vs. $|C_{\FchfPM{1}W^\mp \Fnf{1}}|^2$.\label{sfig:CH1WH2vsCH1WH1:NoDec}]{\includegraphics[width=0.31\textwidth]{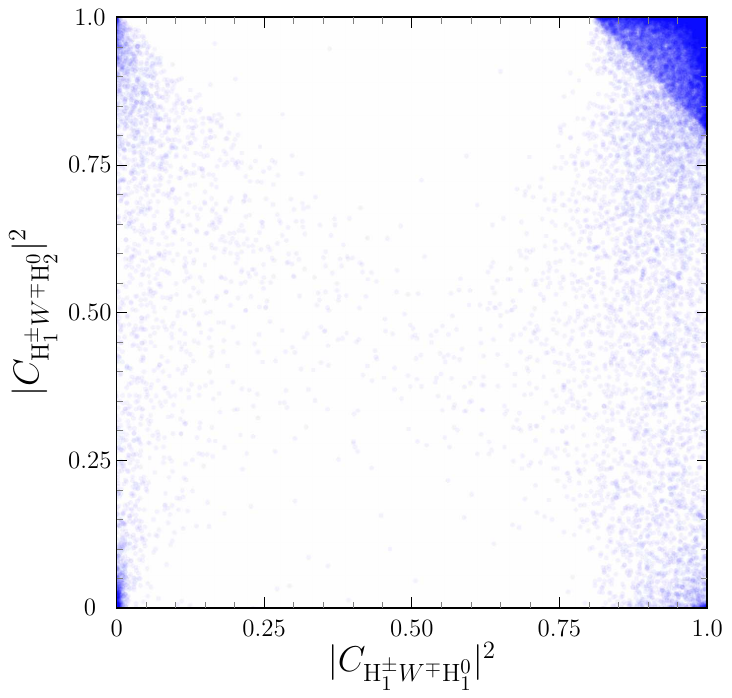}}
\caption{Couplings $C_{\FchfPM{1}W^\mp h}$ and $C_{\FchfPM{1}W^\mp \Fnf{1,2}}$, $M_{\FchfPM{2}}<1$ TeV. \label{fig:CH1:NoDec}} 
\end{figure}

\begin{figure}[!ht]
\centering
\subfloat[$|C_{\FchfPM{1}W^\mp\Fnf{1}}|^2$ vs. $M_{\FchfPM{1}}$.\label{sfig:CH1WH1vsMC1:Dec}]{\includegraphics[width=0.31\textwidth]{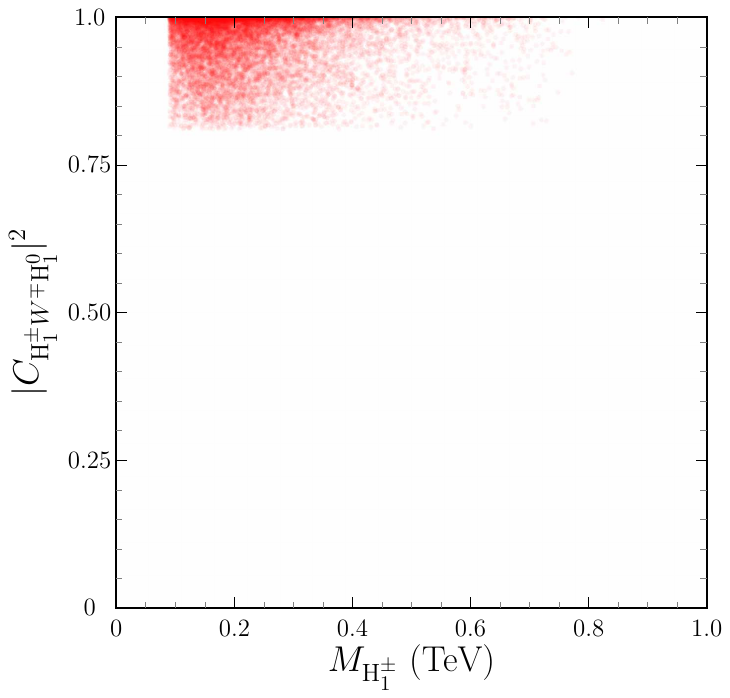}}\quad
\subfloat[$|C_{\FchfPM{1}W^\mp\Fnf{1}}|^2$ vs. $|C_{\FchfPM{1}W^\mp h}|^2$.\label{sfig:CH1WH1vsCH1Wh:Dec}]{\includegraphics[width=0.31\textwidth]{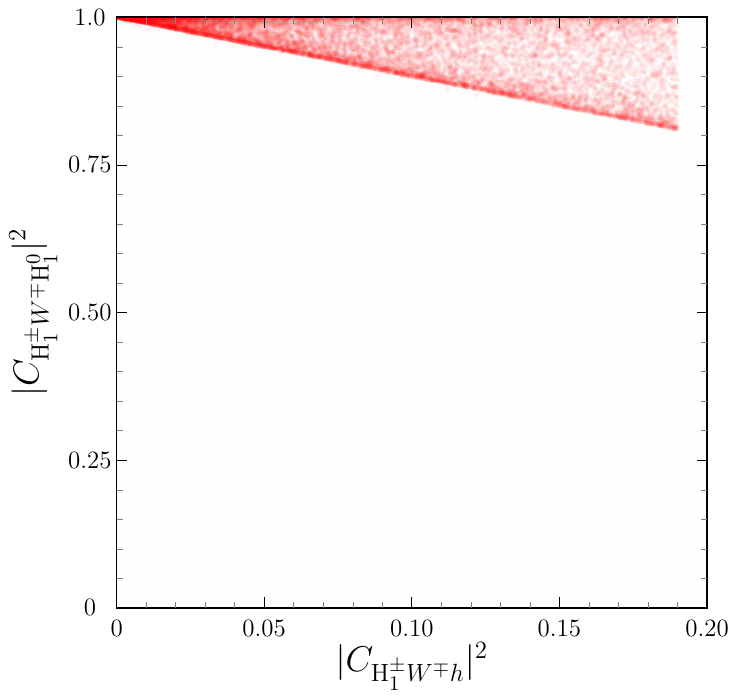}}\quad
\subfloat[$|C_{\FchfPM{1}W^\mp\Fnf{2}}|^2$ vs. $|C_{\FchfPM{1}W^\mp \Fnf{1}}|^2$.\label{sfig:CH1WH2vsCH1WH1:Dec}]{\includegraphics[width=0.31\textwidth]{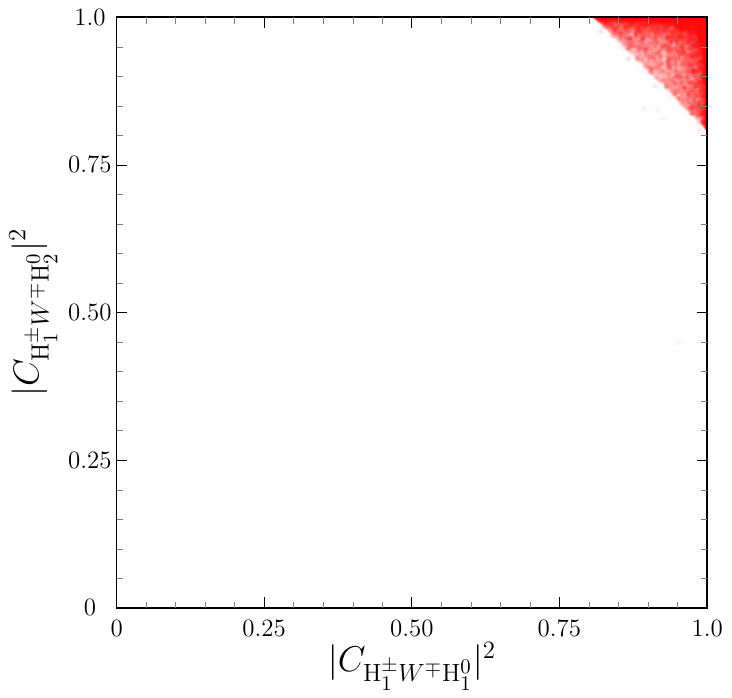}}
\caption{Couplings $C_{\FchfPM{1}W^\mp h}$ and $C_{\FchfPM{1}W^\mp \Fnf{1,2}}$, $M_{\FchfPM{2}}>1$ TeV. \label{fig:CH1:Dec}} 
\end{figure}

Concerning the light charged scalar $\FchfPM{1}$, \cref{fig:CH1:NoDec,fig:CH1:Dec} illustrate properties similar to the ones already discussed for $\Fnf{1}$ and $\Fnf{2}$: (i) allowed ranges of couplings and their independence on $M_{\FchfPM{1}}$, (ii) constraints from unitarity/orthogonality of the mixing matrices $\Umat$ and $\Omat$, and (iii) interplay of different channels, separating again the regimes with $M_{\FchfPM{2}}<1$ TeV (in \cref{fig:CH1:NoDec}) and $M_{\FchfPM{2}}>1$ TeV (in \cref{fig:CH1:Dec}), with the simpler picture arising in the regime with heavy (degenerate) $\FchfPM{2}$, $\Fnf{3}$ and $\Fnf{4}$.
 
\begin{figure}[!ht]
\centering
\subfloat[Max$\left(\Gamma(\Fnf{2}\to X)\right)$ vs. $M_{\Fnf{2}}$.\label{sfig:widths:H}]{\includegraphics[width=0.45\textwidth]{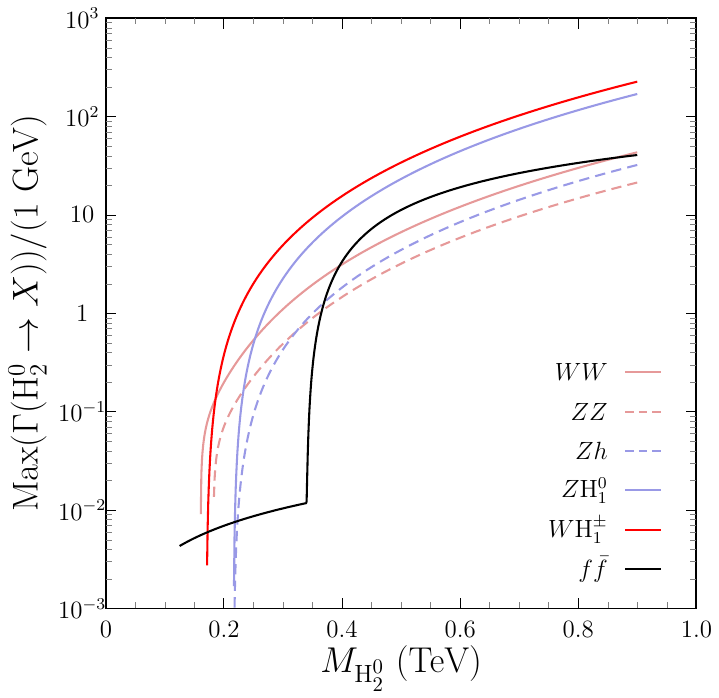}}\qquad
\subfloat[Max$\left(\Gamma(\FchfP{1}\to X)\right)$ vs. $M_{\FchfPM{1}}$.\label{sfig:widths:CH}]{\includegraphics[width=0.45\textwidth]{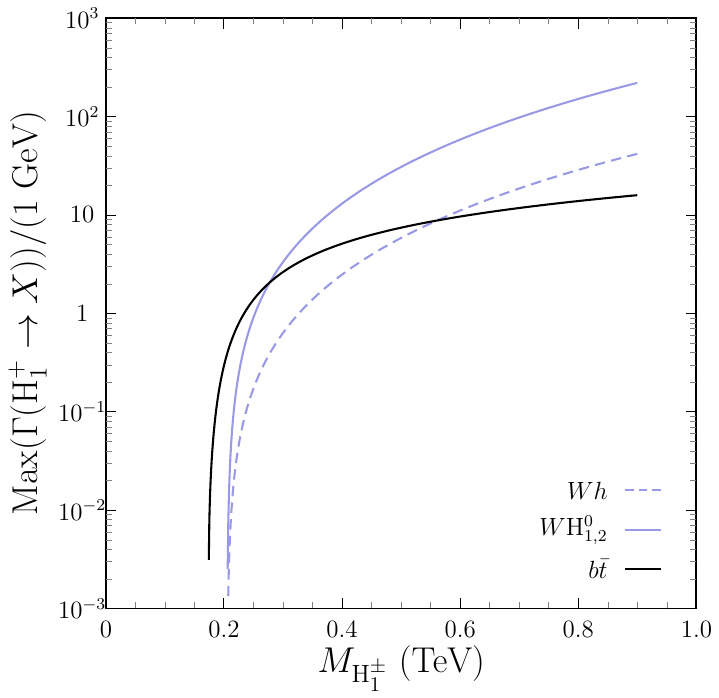}}
\caption{Maximal decay widths of $\Fnf{2}$ and $\FchfPM{1}$. \label{fig:widths}} 
\end{figure}

Collecting these results one can draw a very basic profile of the maximal values of the decay widths of the light scalars\footnote{We represent them for $\Fnf{2}$ rather than for $\Fnf{1}$, since these maximal values coincide, except for $\Gamma(\Fnf{2}\to Z\Fnf{1})$, absent for $\Fnf{1}$ owing to the mass ordering. From the previous discussion it is also clear that different widths might not reach their maximal values simultaneously.} for different channels corresponding to a gauge interaction. As a useful reference, a fermionic decay width into $t\bar t+b\bar b=f\bar f$ is included in the neutral case in \cref{sfig:widths:H} with Yukawa couplings equal to the ones with the SM Higgs. In the charged case in \cref{sfig:widths:CH}, a decay width $\Gamma(\FchfP{1}\to t\bar b)$ deriving from a Yukawa coupling $\frac{1}{\sqrt{2}\vev{}}\left[m_t(1-\gamma_5)-m_b(1+\gamma_5)\right]$ is also included for reference. In \cref{sfig:widths:H,sfig:widths:CH}, when the decay channel involves one of the new scalars --- \eg\ $\Fnf{2}\to Z\Fnf{1}$, $\FchfP{1}\to W^+\Fnf{1}$ ---, the maximal width is computed setting that mass to the lowest allowed value, namely, $m_h$ for $M_{\Fnf{1}}$ and $90$ GeV for $M_{\FchfPM{1}}$.

The bottom line is that even with a scalar potential with a reduced number of parameters, dictated by quartic terms invariant under $A_4$ transformations, and focusing on decays induced by gauge interactions, that is still insufficient to (at least partially) force some pattern in terms of the relevant decay channels.

\clearpage

\section*{Conclusions\label{SEC:Conclusions}}
As discussed in \cite{Miro:2024zka}, real multi-Higgs models with SCPV have new scalars (one charged and two neutral) that must be \emph{light}, \ie\ with masses not much larger than the electroweak scale, if dimensionless quartic couplings respect perturbativity constraints, no matter how large the free mass terms could be. In this work, we have addressed in detail that property in the minimal scenario with 3 Higgs doublets, with the additional requirement of invariance of the quartic part of the potential under some discrete groups, namely, $A_4$ and $\Delta(27)$. We have presented some transparent analytic understanding of that property in \cref{SEC:sym3HDMSCPV}, and further illustrated it with a detailed numerical exploration in \cref{SEC:NumPheno}. Some phenomenological consequences deriving solely from the scalar sector properties have also been presented in \cref{SEC:NumPheno}.

\section*{Acknowledgments}
\textit{Conselleria de Innovación, Universidades, Ciencia y Sociedad Digital} from \textit{Generalitat Valenciana} (Spain) and \emph{Fondo Social Europeo} support JMC, MN, DQ and TT through projects CIDEGENT/2019/024 and CIESGT/2024/21. The authors also acknowledge support from Spanish MICIU/AEI/10.13039/501100011033/ through grant PID2023-151418NB-I00 and the \emph{Severo Ochoa} project CEX2023-001292-S. 

\appendix

\section{Scalar potential and numerical procedure\label{APP:PotNum}}

The numerical exploration of \cref{SEC:NumPheno} is articulated as follows. First, to guarantee that the potential is bounded from below, random values of the quartic parameters are generated and, for fixed $\sum_{a=1}^3|\VEV{\SD{a}}|^2$, it is checked both through a random scan and repeated numerical minimization, that no vev configuration yields $\potV_4(\VEV{\SD{a}})< 0$ (the weaker condition $\potV_4(\VEV{\SD{a}})\leq 0$ is to all effects numerically irrelevant).
At this stage, the overall scale of all quartic couplings does not matter: for simplicity they are all generated uniformly in the range $[-1;1]$.

For a valid bounded-from-below potential, random $\VEV{\SD{a}}$'s and $\qq{23}$ are generated.\footnote{Stationarity conditions are automatically enforced through the non-free quadratic $\qq{1}$, $\qq{2}$, $\qq{3}$, $\qq{12}$ and $\qq{13}$ computed according to \refEqs{eq:Real3HDMA4:Stationarity:v1}-\eqref{eq:Real3HDMA4:Stationarity:th3}.} One can directly generate the vevs satisfying $\vev{1}^2+\vev{2}^2+\vev{3}^2\simeq(246\text{ GeV})^2$ for appropriate electroweak symmetry breaking. For $\qq{23}$, values are generated in a range $[10^{-2};10^{10}]$ GeV$^2$ sufficiently large to allow for the possibility of a (partial) decoupling regime (for better sampling, we generate uniformly $\log_{10}(\qq{23}/\text{GeV}^2)$ in the range $[-2;10]$). 

The squared mass matrices are then computed, and non-negativity of the eigenvalues checked, ensuring that the vev configuration provides a minimum of the potential (the appearance of null eigenvalues, corresponding to the would-be Goldstone bosons, can also be checked within numerical precision). When that is the case, their diagonalization gives the charged and neutral mixing matrices.

Then, to enforce that the lightest neutral scalar has a mass $m_h\simeq 125$ GeV, one can simply take the smallest non-vanishing eigenvalue $m_0^2$ of $\nMM$ and rescale all the parameters of the potential, quadratic and quartic, by a factor $\frac{m_h^2}{m_0^2}$. Stationarity conditions are obviously satisfied all the same, but squared mass matrices are rescaled by $\frac{m_h^2}{m_0^2}$, producing the desired result. One can then finally check if the perturbative unitarity constraints in App.~\ref{APP:2to2} are respected, since the rescaling of all the parameters of the potential can yield values of the quartic parameters (much) larger than the ones in the initial range $[-1;1]$. If they are not respected, that model is of course rejected.

\section{$2\to 2$ scattering and perturbative unitarity constraints\label{APP:2to2}}
We apply the approach of \cite{Ginzburg:2005dt} to obtain the tree-level $2\to 2$ scattering matrices $S_{[I,Y]}$ corresponding to isospin $I=0,2$ and hypercharge $Y=0,1$ initial/final 2-scalar states, both for the $A_4$ invariant quartic potential in \cref{eq:Real3HDMA4:Pot:V4:01} and for the $\Delta(27)$ invariant quartic potential in \cref{eq:Real3HDMD27:Pot:V4:01}. The matrices $S_{[I,Y]}$ have a block-diagonal form for naive, straightforward choices of the basis of scattering states (although that ordering is different in the $A_4$ and $\Delta(27)$ invariant cases). The perturbative unitarity requirement amounts to having the absolute value of all the eigenvalues of the scattering matrices no larger than $16\pi$.

\subsection{One $A_4$ triplet\label{APP:2to2:A4}}
In the $A_4$-invariant case, we have
\begin{equation}\label{eq:2to2:A4:S00:mat}
\begin{aligned}
&S_{[0,0]}=\text{DIAG}\left(S_{[0,0]}^{(1)},S_{[0,0]}^{(2)},S_{[0,0]}^{(2)},S_{[0,0]}^{(2)}\right),\\[5pt]
& S_{[0,0]}^{(1)}=
\begin{pmatrix}
  3\QQ{1}+6\QQ{2} & 2\QQ{1}-2\QQ{2}+\QQ{3} & 2\QQ{1}-2\QQ{2}+\QQ{3}\\
  2\QQ{1}-2\QQ{2}+\QQ{3} & 3\QQ{1}+6\QQ{2} & 2\QQ{1}-2\QQ{2}+\QQ{3}\\
  2\QQ{1}-2\QQ{2}+\QQ{3} & 2\QQ{1}-2\QQ{2}+\QQ{3} & 3\QQ{1}+6\QQ{2}
\end{pmatrix},\\[5pt]
& S_{[0,0]}^{(2)}=
\begin{pmatrix}
 \QQ{1}-\QQ{2}+2\QQ{3} & 3\QQ{4}\\
 3\QQ{4}^\ast & \QQ{1}-\QQ{2}+2\QQ{3}
\end{pmatrix}.
\end{aligned}
\end{equation}
The eigenvalues of $S_{[0,0]}^{(1)}$, $S_{[0,0]}^{(2)}$ are
\begin{equation}\label{eq:2to2:A4:S00:eig}
\begin{aligned}
&\text{Eig}(S_{[0,0]}^{(1)})=\left\{\QQ{1}+8\QQ{2}-\QQ{3},\QQ{1}+8\QQ{2}-\QQ{3},7\QQ{1}+2\QQ{2}+2\QQ{3}\right\},\\
&\text{Eig}(S_{[0,0]}^{(2)})=\left\{\QQ{1}-\QQ{2}+2\QQ{3}+3\abs{\QQ{4}},\QQ{1}-\QQ{2}+2\QQ{3}-3\abs{\QQ{4}}\right\}.
\end{aligned}
\end{equation}
We also have
\begin{equation}\label{eq:2to2:A4:S01:mat}
\begin{aligned}
&S_{[0,1]}=\text{DIAG}\left(S_{[0,1]}^{(1)},S_{[0,1]}^{(2)},S_{[0,1]}^{(2)},S_{[0,1]}^{(2)}\right),\\[5pt] 
& S_{[0,1]}^{(1)}=
\begin{pmatrix}
  \QQ{1}+2\QQ{2} & \QQ{3} & \QQ{3}\\
  \QQ{3} & \QQ{1}+2\QQ{2} & \QQ{3}\\
  \QQ{3} & \QQ{3} & \QQ{1}+2\QQ{2}
\end{pmatrix},\qquad
 S_{[0,1]}^{(2)}=
\begin{pmatrix}
 \QQ{1}-\QQ{2} & \QQ{4}\\
 \QQ{4}^\ast & \QQ{1}-\QQ{2}
\end{pmatrix},
\end{aligned}
\end{equation}
with corresponding eigenvalues given by
\begin{equation}\label{eq:2to2:A4:S01:eig}
\begin{aligned}
&\text{Eig}(S_{[0,1]}^{(1)})=\left\{\QQ{1}+2\QQ{2}-\QQ{3}, \QQ{1}+2\QQ{2}-\QQ{3}, \QQ{1}+2\QQ{2}+2\QQ{3}\right\},\\
&\text{Eig}(S_{[0,1]}^{(2)})=\left\{\QQ{1}-\QQ{2}+\abs{\QQ{4}},\QQ{1}-\QQ{2}-\abs{\QQ{4}}\right\}.
\end{aligned}
\end{equation}
The matrix $S_{[2,0]}$ is proportional to the identity,
\begin{equation}\label{eq:2to2:A4:S20:mat}
 S_{[2,0]}=(\QQ{1}-\QQ{2}+\QQ{3})\mathbf{1}_3\,.
\end{equation}
Finally,
\begin{equation}\label{eq:2to2:A4:S21:mat}
\begin{aligned}
&S_{[2,1]}=\text{DIAG}\left(S_{[2,1]}^{(1)},S_{[2,1]}^{(2)}\right),\\[5pt]
&S_{[2,1]}^{(1)}=
\begin{pmatrix}
  \QQ{1}+2\QQ{2} & \QQ{4}^\ast & \QQ{4}^\ast\\
  \QQ{4} & \QQ{1}+2\QQ{2} & \QQ{4}^\ast\\
  \QQ{4} & \QQ{4} & \QQ{1}+2\QQ{2}
\end{pmatrix},\qquad
S_{[2,1]}^{(2)}=(\QQ{1}-\QQ{2}+\QQ{3})\mathbf{1}_3\,.
\end{aligned}
\end{equation}
The matrix $S_{[2,1]}^{(1)}$ above corresponds to the one triplet $A_4$-invariant case, with $\QQ{4}\in\mathbb{C}$; as mentioned previously, requiring in addition CP invariance, $\QQ{4}\in\mathbb{R}$, in which case the eigenvalues of $S_{[2,1]}^{(1)}$ are
\begin{equation}\label{eq:2to2:A4:S21:eig}
 \text{Eig}(S_{[2,1]}^{(1)})=\left\{\QQ{1}+2\QQ{2}+2\QQ{4},\QQ{1}+2\QQ{2}-\QQ{4},\QQ{1}+2\QQ{2}-\QQ{4}\right\}.
\end{equation}

\subsection{One $\Delta(27)$ triplet\label{APP:2to2:Delta27}}
In the $\Delta(27)$-invariant case, we show the scattering matrices with $\QQ{\Delta}\in\mathbb{C}$; requiring CP invariance we just need to set $\im{\QQ{\Delta}}=0$. We have
\begin{equation}\label{eq:2to2:D27:S00:mat}
\begin{aligned}
&S_{[0,0]}=\text{DIAG}\left(S_{[0,0]}^{(1)},S_{[0,0]}^{(2)},S_{[0,0]}^{(2)}\right),\\[5pt]
& S_{[0,0]}^{(1)}=
\begin{pmatrix}
 3\QQ{1}+6\QQ{2}+6\QQ{3} & 4\QQ{2}+2\QQ{3} & 4\QQ{2}+2\QQ{3}\\
 4\QQ{2}+2\QQ{3} & 3\QQ{1}+6\QQ{2}+6\QQ{3} & 4\QQ{2}+2\QQ{3}\\
 4\QQ{2}+2\QQ{3} & 4\QQ{2}+2\QQ{3} & 3\QQ{1}+6\QQ{2}+6\QQ{3}
\end{pmatrix},\\[5pt]
& S_{[0,0]}^{(2)}=
\begin{pmatrix}
  2\QQ{2}+4\QQ{3} & \frac{3}{2}\QQ{\Delta}^\ast & \frac{3}{2}\QQ{\Delta}\\
  \frac{3}{2}\QQ{\Delta} & 2\QQ{2}+4\QQ{3} & \frac{3}{2}\QQ{\Delta}^\ast\\
  \frac{3}{2}\QQ{\Delta}^\ast & \frac{3}{2}\QQ{\Delta}^\ast & 2\QQ{2}+4\QQ{3}
\end{pmatrix}.
\end{aligned}
\end{equation}
The eigenvalues of $S_{[0,0]}^{(1)}$ and $S_{[0,0]}^{(2)}$ are
\begin{equation}\label{eq:2to2:D27:S00:eig}
\begin{aligned}
&\text{Eig}(S_{[0,0]}^{(1)})=\left\{3\QQ{1}+2\QQ{2}+4\QQ{3},3\QQ{1}+2\QQ{2}+4\QQ{3},3\QQ{1}+14\QQ{2}+10\QQ{3}\right\}\,,\\
&\text{Eig}(S_{[0,0]}^{(2)})=\left\{2\QQ{2}+4\QQ{3}+3\re{\QQ{\Delta}},2\QQ{2}+4\QQ{3}-\frac{3}{2}\re{\QQ{\Delta}}\pm\frac{3\sqrt{3}}{2}\im{\QQ{\Delta}}\right\}\,.
\end{aligned}
\end{equation}
On the other hand,
\begin{equation}\label{eq:2to2:D27:S01:mat}
\begin{aligned}
&S_{[0,1]}=\text{DIAG}\left(S_{[0,1]}^{(1)},S_{[0,1]}^{(2)},S_{[0,1]}^{(2)}\right),\\[5pt] 
& S_{[0,1]}^{(1)}=
\begin{pmatrix}
 \QQ{1}+2\QQ{2}+2\QQ{3} & 2\QQ{3} & 2\QQ{3}\\
 2\QQ{3} & \QQ{1}+2\QQ{2}+2\QQ{3} & 2\QQ{3}\\
 2\QQ{3} & 2\QQ{3} & \QQ{1}+2\QQ{2}+2\QQ{3}
\end{pmatrix},
\\[5pt]
& S_{[0,1]}^{(2)}=
\begin{pmatrix}
  2\QQ{2} & \frac{\QQ{\Delta}^\ast}{2} & \frac{\QQ{\Delta}}{2}\\
  \frac{\QQ{\Delta}}{2} & 2\QQ{2} & \frac{\QQ{\Delta}^\ast}{2}\\
  \frac{\QQ{\Delta}^\ast}{2} & \frac{\QQ{\Delta}}{2} & 2\QQ{2}
\end{pmatrix}.
\end{aligned}
\end{equation}\label{eq:2to2:D27:S01:eig}
The corresponding eigenvalues are
\begin{equation}
\begin{aligned}
&\text{Eig}(S_{[0,1]}^{(1)})=\left\{\QQ{1}+2\QQ{2},\QQ{1}+2\QQ{2},\QQ{1}+2\QQ{2}+6\QQ{3}\right\},\\
&\text{Eig}(S_{[0,1]}^{(2)})=\left\{2\QQ{2}+\re{\QQ{\Delta}},2\QQ{2}-\frac{1}{2}\re{\QQ{\Delta}}\pm\frac{\sqrt 3}{2}\im{\QQ{\Delta}}\right\}.
\end{aligned}
\end{equation}
$S_{[2,0]}$ is, as in the $A_4$-invariant case in \refEq{eq:2to2:A4:S20:mat}, proportional to the identity matrix,
\begin{equation}\label{eq:2to2:D27:S20:mat}
 S_{[2,0]}=(2\QQ{2}+2\QQ{3})\mathbf{1}_3\,.
\end{equation}
Finally,
\begin{equation}\label{eq:2to2:D27:S21:mat}
\begin{aligned}
&S_{[2,1]}=\text{DIAG}\left(S_{[2,1]}^{(1)},S_{[2,1]}^{(1)},S_{[2,1]}^{(1)}\right),\\[5pt]
&S_{[2,1]}^{(1)}=
\begin{pmatrix}
  \QQ{1}+2\QQ{2}+2\QQ{3}& \frac{\QQ{\Delta}}{\sqrt 2}\\
  \frac{\QQ{\Delta}^\ast}{\sqrt 2} & 2\QQ{2}+2\QQ{3}
\end{pmatrix},
\end{aligned}
\end{equation}
and the eigenvalues of $S_{[2,1]}^{(1)}$ are
\begin{equation}\label{eq:2to2:D27:S21:eig}
 \text{Eig}(S_{[2,1]}^{(1)})=\left\{\frac{1}{2}\left(\QQ{1}+4\QQ{2}+4\QQ{3}\pm\sqrt{\QQ{1}^2+2\abs{\QQ{\Delta}}^2}\right)\right\}.
\end{equation}

\clearpage

\bibliographystyle{apsrev4-2}
\bibliography{final.bib}

\end{document}